\documentclass[prd,aps,10pt,twocolumn,nofootinbib,preprintnumbers,floatfix]{revtex4}

\usepackage{graphicx}
\usepackage[hang,tight,raggedright]{subfigure}
\usepackage[hyperfootnotes=false,bookmarks=false]{hyperref}
\usepackage{amsmath}

\newcommand{\eq}[1]{Eq.~\eqref{eq:#1}}
\newcommand{\eqs}[2]{Eqs.~\eqref{eq:#1} and \eqref{eq:#2}}
\renewcommand{\sec}[1]{Sec.~\ref{sec:#1}}

\newcommand{\subsec}[1]{Sec.~\ref{subsec:#1}}
\newcommand{\fig}[1]{Fig.~\ref{fig:#1}}
\newcommand{\figs}[2]{Figs.~\ref{fig:#1} and \ref{fig:#2}}

\newcommand{\abs}[1]{\lvert#1\rvert}
\newcommand{\ord}[1]{\mathcal{O}(#1)}

\newcommand{\df}{\mathrm{d}}

\newcommand{\GeV}{\,\mathrm{GeV}}
\newcommand{\TeV}{\,\mathrm{TeV}}

\newcommand{\nn}{\nonumber}

\newcommand{\cut}{\mathrm{cut}}

\newcommand{\LO}{\mathrm{LO}}

\newcommand{\pT}{p_{THjj}}
\newcommand{\dphi}{\Delta\phi_{H\!-\!jj}}

\begin{document}


\preprint{\vbox{\hbox{DESY 13-029}}}

\title{NLO Uncertainties in Higgs + 2 Jets From Gluon Fusion}

\author{Shireen Gangal}
\author{Frank J.~Tackmann\vspace{0.5ex}}

\affiliation{Theory Group, Deutsches Elektronen-Synchrotron (DESY), D-22607 Hamburg, Germany\vspace{2ex}}

\date{February 21, 2013}

\begin{abstract}
A central ingredient in establishing the properties of the newly discovered Higgs-like boson is the isolation of its production via vector boson fusion (VBF). With the typical experimental selection cuts, the VBF sample is contaminated by a $\sim 25\%$ fraction from Higgs + 2 jet production via gluon fusion (ggF), which has large perturbative uncertainties. We perform a detailed study of the perturbative uncertainties in the NLO predictions for $pp\to H+2$ jets via ggF used by the ATLAS and CMS Collaborations, with the VBF selection cuts of their current $H\to\gamma\gamma$ analyses. We discuss in detail the application of the so-called ``ST method'' for estimating fixed-order perturbative uncertainties in this case and also consider generalizations of it. Qualitatively, our results apply equally to other decay channels with similar VBF selection cuts. Typical VBF selections include indirect restrictions or explicit vetoes on additional jet activity, primarily to reduce non-Higgs backgrounds. We find that such restrictions have to be chosen carefully and are not necessarily beneficial for the purpose of distinguishing between the VBF and ggF production modes, since a modest reduction in the relative ggF contamination can be easily overwhelmed by its quickly increasing perturbative uncertainties.
\end{abstract}

\maketitle

\section{Introduction}
\label{sec:intro}

With the discovery of a Higgs-like boson by the ATLAS and CMS Collaborations~\cite{:2012gk,:2012gu}, a central ingredient in measuring the properties of the new particle is to separate out the different production mechanisms via gluon-gluon fusion (ggF) and vector boson fusion (VBF).

To maximize the signal sensitivity, the experimental analyses separate the data into various exclusive selection categories, based on the number of jets (``jet bins'') and other criteria. One selection category, designed to isolate a clean VBF signal, is the production in connection with two jets that are widely separated in rapidity. To further enhance the VBF signal and reduce non-Higgs backgrounds as well as the sizable contamination from ggF production, additional kinematic selection criteria are applied. A characteristic feature of the VBF process is that it is accompanied by few extra gluon emissions, because of its color structure and incoming quarks. The same is not the case for ggF production or generic non-Higgs backgrounds, so the VBF signal tends to be most significant in the exclusive 2-jet region of phase space with two forward jets and little additional radiation. Therefore, when optimizing its significance, whether in a cut-based approach or via multivariate techniques, one dominantly selects events from this region.

In general, placing a restriction on additional real emissions induces Sudakov logarithms at each order in the perturbative series. In the limit of very tight restrictions, the logarithms become large and must be resummed to all orders in the strong coupling constant $\alpha_s$ to obtain a meaningful perturbative prediction. Often, the experimentally relevant region is an intermediate one, where the logarithmic corrections are already sizable but their resummation is not yet strictly necessary and a fixed-order expansion can still be applied. In this case, however, it is important that the possible sizable effects of higher-order logarithms are reflected in the perturbative uncertainty estimate for the fixed-order prediction.

Due to the incoming gluons in ggF and the associated large color factor, the logarithmic corrections in this intermediate region are indeed sizable. This was shown explicitly for the 0-jet bin in Refs.~\cite{Berger:2010xi, Stewart:2011cf}, which has been computed and studied extensively to NNLO~\cite{Catani:2001cr, Anastasiou:2004xq, Anastasiou:2007mz, Grazzini:2008tf, Anastasiou:2008ik, Anastasiou:2009bt}. A numerically important ggF contribution to the VBF-like 2-jet selection is the partonic $gg\to Hgg$ process, where both incoming and outgoing gluons generate additional radiation. When restricting that radiation by forcing the kinematics into the exclusive 2-jet region, the logarithmic corrections can be expected to be at least as large and likely larger than in the 0-jet case. Hence, the perturbative uncertainties have to estimated carefully, in particular since here NNLO predictions are not available.

In Ref.~\cite{Stewart:2011cf} a simple method was devised that explicitly takes into account the size of the logarithmic corrections in the fixed-order uncertainty estimate (for which a simple scale variation in the exclusive jet cross section is insufficient). It has been adopted in Refs.~\cite{HiggsCombinationNote, Dittmaier:2012vm}, and is being employed in various exclusive analyses at the LHC and the Tevatron. It is sometimes referred to as the ``ST method.'' Here, we apply this method to provide robust uncertainty estimates for the NLO calculation~\cite{Campbell:2006xx, Campbell:2010cz} in the exclusive 2-jet bin in ggF production that is currently being used in the ATLAS and CMS Higgs analyses. An independent NLO calculation has been performed recently in Ref.~\cite{vanDeurzen:2013rv}. We also discuss the application to more general cuts restricting to the exclusive $2$-jet region. The application to multivariate selection techniques will be discussed in a forthcoming publication~\cite{toappear}.

In principle, by performing a higher-order logarithmic resummation, one can gain additional information, which allows for refined perturbative predictions and uncertainty estimates, see e.g. Refs.~\cite{Berger:2010xi, Banfi:2012yh, Becher:2012qa, Tackmann:2012bt, Banfi:2012jm, Liu:2012sz, Jouttenus:2013hs} for recent applications to Higgs + 0 or 1 jets.
Although technically more demanding, similar methods could be used in the future to provide improved predictions for the exclusive $gg\to H+2 $ jets cross section.
An important step in this direction was made in Refs.~\cite{Campbell:2012am, Frederix:2012ps}, where the fixed NLO predictions for $gg\to H+2$ jets are matched to a parton shower, which on top of the NLO corrections provides a leading-logarithmic (LL) resummation in the exclusive 2-jet region. Here, our NLO uncertainty analysis provides an important baseline for future studies, as it is often difficult to obtain a robust resummed uncertainty estimate from a LL resummation alone. 
For Higgs + 2 jets the all-order resummation of soft gluon emissions in the presence of a central jet veto has also been performed~\cite{Kidonakis:1998nf, Forshaw:2007vb}.

In the next section, we give a general discussion of jet binning uncertainties, reviewing and extending the method to estimate fixed-order uncertainties introduced in Ref.~\cite{Stewart:2011cf}. In \sec{application} we discuss and validate its application to the ggF contribution in the VBF-like 2-jet selection. In \sec{results} we present our results for the NLO uncertainty estimates, implementing the VBF selection cuts used in the $H\to\gamma\gamma$ analyses by the ATLAS~\cite{ATLAS:2012goa, ATLAS-CONF-2012-168} and CMS~\cite{CMS:2012zwa} experiments. Our findings apply equally to other decay channels where similar VBF selection cuts are applied. In \sec{results} we also discuss the implications of the ggF uncertainties for the VBF-ggF separation. We conclude in \sec{conclusions}.

\section{Jet Binning Uncertainties}
\label{sec:jetbinreview}

Consider the \emph{inclusive} $N$-jet cross section, $\sigma_{\geq N}$, for some process containing at least $N$ jets. We will assume that $\sigma_{\geq N}$ is a sufficiently inclusive quantity such that it can be computed in fixed-order perturbation theory. We are interested in the case where $\sigma_{\geq N}$ is divided up into a corresponding \emph{exclusive} $N$-jet cross section, $\sigma_N$, and a remainder $\sigma_{\geq N+1}$,
\begin{equation} \label{eq:sigmaNgeneral}
\sigma_{\geq N}
= \sigma_{N}(\text{excl. cut}) + \sigma_{\geq N+1}(\text{inverse excl. cut})
\,.\end{equation}
All three cross sections here have the \emph{same} selection cuts applied that identify the leading $N$ signal jets. What defines $\sigma_N$ to be ``exclusive'' is that the additional exclusive cut applied to it restricts the phase space of additional emissions in such a way that $\sigma_N$ is dominated by configurations close to the $N$-parton Born kinematics. In particular, at leading order (LO) in perturbation theory $\sigma_{\geq N}^\LO = \sigma_N^\LO$, while relative to these, $\sigma_{\geq N+1}$ is suppressed by $\ord{\alpha_s}$. In other words, $\sigma_{\geq N+1}$ requires at least one additional emission to be nonvanishing. Hence, we can consider it an inclusive $(N+1)$-jet cross section with at least $N+1$ jets.%
\footnote{It should be stressed that $\sigma_{\geq N+1}$ here is defined by inverting the exclusive cut that defines $\sigma_N$ and so does not necessarily require the explicit identification of another well-separated jet via a jet algorithm. The variables we will use, $\pT$ and $\dphi$ are examples of this.}

In the simplest case, $\sigma_{\geq N}$ is divided into the two jet bins $\sigma_N$ and $\sigma_{\geq N+1}$ by using a cut on some kinematic variable, $p_{N+1}$, which characterizes additional emissions, with $p_{N+1} = 0$ for a tree-level $N$-parton state. Typical examples would be the $p_T$ of the $N+1$st jet or the total $\abs{\vec{p}_T}$ of the underlying $N$-jet system. The two jet bins then correspond to the integrals of the differential spectrum $\df\sigma/\df p_{N+1}$ above and below some cut,
\begin{align} \label{eq:pcut}
\sigma_{\geq N} &= \int_0^{p^\cut}\!\df p_{N+1}\, \frac{\df\sigma_{\geq N}}{\df p_{N+1}} + \int_{p^\cut}\!\df p_{N+1}\,
\frac{\df\sigma_{\geq N}}{\df p_{N+1}}
\nn\\
&\equiv  \sigma_N(p^\cut) + \sigma_{\geq N+1}(p^\cut)
\,.\end{align}
In general, the jet bin boundary could be a much more complicated function of phase space, for example in a multivariate analysis.
In \sec{results}, we will also consider the next-to-simplest case of a two-dimensional rectangular cut on two kinematic variables.

We are interested in the uncertainties involved in the binning. The covariance matrix for $\{\sigma_N, \sigma_{\geq N+1}\}$ is a symmetric $2\times 2$ matrix with three independent parameters. A convenient and general parametrization is to write it in terms of two components,
\begin{align} \label{eq:Cgeneral}
C &=
\begin{pmatrix}
(\Delta^{\rm y}_{N})^2 &  \Delta^{\rm y}_{N}\,\Delta^{\rm y}_{\geq N+1}  \\
\Delta^{\rm y}_{N}\,\Delta^{\rm y}_{\geq N+1} & (\Delta^{\rm y}_{\geq N+1})^2
\end{pmatrix}
+
\begin{pmatrix}
 \Delta_\cut^2 &  - \Delta_\cut^2 \\
-\Delta_\cut^2 & \Delta_\cut^2
\end{pmatrix}
.\end{align}
Here, the first term is an absolute ``yield'' uncertainty, denoted with a superscript ``y,'' which (by definition) is 100\% correlated between the two bins $\sigma_N$ and $\sigma_{\geq N+1}$. The second term is a ``migration'' uncertainty between the bins and corresponds to the uncertainty introduced by the binning cut. It has the same absolute size, $\Delta_\cut$, for both bins and is 100\% anticorrelated between them, such that it drops out when the two bins are added. Hence, the total uncertainty for each bin is given by
\begin{align} \label{eq:DeltaN}
\Delta_{N}^2 &= (\Delta_N^{\rm y})^2 + \Delta_\cut^2
\nn\\
\Delta_{\geq N+1}^2 &= (\Delta_{\geq N+1}^{\rm y})^2 + \Delta_\cut^2
\,,\end{align}
while the total uncertainty on their sum, i.e., on $\sigma_{\geq N}$, is given by the total yield uncertainty,
\begin{equation} \label{eq:DeltageqN}
\Delta_{\geq N} = \Delta^{\rm y}_{\geq N} = \Delta^{\rm y}_N + \Delta^{\rm y}_{\geq N+1}
\,.\end{equation}

Considering the perturbative uncertainties, the basic question is how each of the uncertainties in \eq{Cgeneral} can be evaluated. The fixed-order prediction provides us with two independent pieces of information, namely the variations obtained by the standard scale variations, which we denote as $\Delta^\mu_{\geq N}$, $\Delta^\mu_{N}$, $\Delta^\mu_{\geq N+1}$, and which satisfy $\Delta^\mu_{\geq N} = \Delta^\mu_{N} + \Delta^\mu_{\geq N+1}$. Following Ref.~\cite{Stewart:2011cf}, we start by assuming that the standard fixed-order scale variations can be used to obtain a reliable estimate of the total uncertainties in the \emph{inclusive} cross sections (which is of course the common assumption underlying any inclusive fixed-order calculation). Hence, we impose the two well-motivated boundary conditions,
\begin{equation}
\Delta_{\geq N} = \Delta^\mu_{\geq N}
\,,\qquad
\Delta_{\geq N+1} = \Delta^\mu_{\geq N+1}
\,.\end{equation}
Together with \eqs{DeltaN}{DeltageqN}, these lead to
\begin{align} \label{eq:conditions}
\text{(i)} && \Delta^\mu_{\geq N} &= \Delta^{\rm y}_N + \Delta^{\rm y}_{\geq N+1}
\,,\nn\\
\text{(ii)} && (\Delta^\mu_{\geq N+1})^2 &= (\Delta_{\geq N+1}^{\rm y})^2 + \Delta_\cut^2
\,.\end{align}
Thus, the question is how to divide up $\Delta^\mu_{\geq N+1}$ between $\Delta_{\geq N+1}^{\rm y}$ and $\Delta_\cut$ in order to satisfy condition (ii). Condition (i) then determines $\Delta_N^{\rm y}$. The nontrivial effect $\Delta_\cut$ can have is on the size of $\Delta_N$ as well as on the off-diagonal entries in \eq{Cgeneral}, which determine the correlation between $\Delta_N$ and $\Delta_{\geq N+1}$.

Clearly, the simplest is to neglect the effect of $\Delta_\cut$ altogether and to directly use common scale variations to estimate the uncertainties, i.e., to take
\begin{align} \label{eq:direct}
\Delta_N^{\rm y} &= \Delta_{\geq N}^\mu - \Delta_{\geq N + 1}^\mu \equiv \Delta_N^\mu
\,,\quad \Delta_{\geq N + 1}^{\rm y} = \Delta_{\geq N + 1}^\mu
\,,\nn\\
\Delta_\cut &= 0
\,,\end{align}
which leads to
\begin{equation} \label{eq:Cdirect}
\mspace{-40mu}\text{direct:} \mspace{40mu}
C =
\begin{pmatrix}
(\Delta^\mu_{N})^2 &  \Delta^\mu_{N}\,\Delta^\mu_{\geq N+1}  \\
\Delta^\mu_{N}\,\Delta^\mu_{\geq N+1} & (\Delta^\mu_{\geq N+1})^2
\end{pmatrix}
.\end{equation}
Note that since $\sigma_{\geq N+1}$ starts at higher order in perturbation theory than $\sigma_{\geq N}$, its relative uncertainty $\Delta_{\geq N+1}^\mu/\sigma_{\geq N+1}$ will typically be (much) larger than $\sigma_{\geq N}$'s relative uncertainty $\Delta_{\geq N}^\mu/\sigma_{\geq N}$. This means one cannot simply apply the latter as the relative yield uncertainty in each bin by taking $\Delta_i^{\rm y} = (\Delta_{\geq N}^\mu/\sigma_{\geq N}) \sigma_i$, as this would violate the condition $\Delta_{\geq N+1} = \Delta_{\geq N+1}^\mu$. This point has already been emphasized in earlier studies~\cite{Anastasiou:2009bt}.

The direct scale variation choice is reasonable as long as the effect of $\Delta_\cut$ is indeed negligible. It is certainly justified if numerically $\Delta^\mu_{\geq N} \gg \Delta^\mu_{\geq N+1}$, since any uncertainty due to migration effects can be, at most as large as $\Delta^\mu_{\geq N+1}$ [by virtue of condition (ii)]. This can happen, for example, when $\Delta^\mu_{\geq N}$ is sizable due to large perturbative corrections in $\sigma_{\geq N}$ and/or the binning cut is very loose (i.e., is cutting out only a small fraction of phase space) such that $\sigma_{\geq N+1}$ is numerically small to begin with.

In perturbation theory, the effect of the binning cut is to introduce Sudakov double logarithms in the perturbative series of $\sigma_N$ and $\sigma_{\geq N+1}$, which have opposite sign and cancel in the sum of the two bins, schematically,
\begin{align}
\sigma_{\geq N} &\simeq \sigma_B [1 + \alpha_s + \alpha_s^2 + \ord{\alpha_s^3} \bigr]
\,,\nn\\
\sigma_{\geq N+1} &\simeq \sigma_B \bigl[\alpha_s (L^2 + L + 1)
\nn\\ & \quad
+ \alpha_s^2 (L^4 + L^3 + L^2 + L + 1) + \ord{\alpha_s^3 L^6} \bigr]
\,,\nn\\
\sigma_{N} &= \sigma_{\geq N} - \sigma_{\geq N+1}
\,,\end{align}
where $\sigma_B$ denotes the Born cross section and $L$ is a Sudakov logarithm, e.g., for \eq{pcut}, $L = \ln(p^\cut/Q)$, where $Q\sim m_H$ is a typical hard scale. As the binning cut becomes tighter ($p^\cut$ becomes smaller) the logarithms grow in size. Once the logarithms are $\ord{1}$ numbers, one is in the transition region and the logarithms will start to dominate the perturbative series of $\sigma_{\geq N+1}$, and there will be sizable cancellations in $\sigma_N$ between the perturbative series for $\sigma_{\geq N}$ and the logarithmic series in $\sigma_{\geq N+1}$. Eventually, the logarithms will grow large enough to overcome the $\alpha_s$ suppression and $\sigma_{N}$ becomes negative, at which point one is in the resummation region and the fixed-order expansion has broken down.

The perturbative migration uncertainty $\Delta_\cut$ can be directly associated with the perturbative uncertainty in the logarithmic series induced by the binning, and so should not be neglected once the logarithms have a noticeable effect. In particular, as demonstrated in Ref.~\cite{Stewart:2011cf}, the simple choice in \eqs{direct}{Cdirect} can easily lead to an underestimate of $\Delta_N$ in the region where there are sizable numerical cancellations between the two series in $\sigma_{\geq N}$ and $\sigma_{\geq N+1}$. Since in this region the dominant contribution to $\sigma_{\geq N+1}$ comes from the logarithmic series, varying the scales in $\sigma_{\geq N+1}$ directly tracks the size of the logarithms, which means we can use $\Delta_\cut = \Delta^\mu_{\geq N+1}$ as an estimate for the binning uncertainty, which is the basic idea of Ref.~\cite{Stewart:2011cf}. From \eq{conditions}, we then find
\begin{align} \label{eq:ST}
\Delta_N^{\rm y} &= \Delta_{\geq N}^\mu
\,,\qquad \Delta_{\geq N + 1}^{\rm y} = 0
\,,\nn\\
\Delta_\cut &= \Delta_{\geq N+1}^\mu
\,,\end{align}
such that
\begin{equation} \label{eq:CST}
\text{ST:} \quad
C =
\begin{pmatrix}
(\Delta^\mu_{\geq N})^2 + (\Delta^\mu_{\geq N+1})^2 &  - (\Delta^\mu_{\geq N+1})^2 \\
- (\Delta^\mu_{\geq N+1})^2 & (\Delta^\mu_{\geq N+1})^2
\end{pmatrix}
.\end{equation}
Since $\Delta_{\geq N+1}^\mu$ is now used as $\Delta_\cut$, the effective outcome is that one treats $\Delta^\mu_{\geq N}$ and $\Delta^\mu_{\geq N+1}$ as uncorrelated.

\begin{table*}[t!]
\begin{tabular}{c|c|c|c}
\hline\hline
& ATLAS  &  CMS loose & CMS tight  \\
\hline
& anti-$k_T$ $R = 0.4$ & anti-$k_T$ $R = 0.5$ & anti-$k_T$ $R = 0.5$
\\
2-jet selection & $p_{Tj}\!>\! 25\GeV$ for $\abs{\eta_{j}}\! <\! 2.5$ & jet 1: $ p_{Tj}\!>\! 30\GeV$, $\abs{\eta_{j}}\!<\! 4.7$
& $p_{Tj}\!>\! 30\GeV$, $\abs{\eta_{j}} \!<\! 4.7 $
\\
& $p_{Tj}\!>\! 30\GeV$ for $2.5\! <\! \abs{\eta_{j}} \!<\! 4.5 $ & jet 2: $p_{Tj}\!>\! 20\GeV$,  $\abs{\eta_{j}}\!<\! 4.7$  &
\\\hline
$\Delta\eta_{jj} = \abs{\eta_{j1} - \eta_{j2}}$ & $ > 2.8$  & $ > 3.0 $ & $ > 3.0 $
\\
$m_{jj}$ &  $> 400\GeV$ & $> 250\GeV$ & $ >500\GeV$
\\
$|\eta_{H} - (\eta_{j1} + \eta_{j2})/2|$ & -  & $ < 2.5 $ & $<2.5 $
\\\hline
$\Delta\phi_{H\!-\!jj}$ & $>  2.6 $ & $>  2.6 $ & $>  2.6 $\\
\hline\hline
\end{tabular}
\caption{VBF selection cuts we use, corresponding to the $H\to\gamma\gamma$ analyses by ATLAS~\cite{ATLAS:2012goa, ATLAS-CONF-2012-168} and CMS~\cite{CMS:2012zwa}. CMS loose excludes events that pass CMS tight. The cut on $\dphi$ in the last row
is treated specially as an exclusive binning cut.}
\label{tab:cuts}
\end{table*}

More generally, we can introduce a parameter $0 \leq \rho \leq 1$, which controls the fraction of $\Delta_{\geq N+1}^\mu$ assigned to $\Delta^{\rm y}_{\geq N+1}$, such that
\begin{align}
\Delta^{\rm y}_{N} &= \Delta_{\geq N}^\mu - \rho\, \Delta_{\geq N+1}^\mu
\,,\quad
\Delta^{\rm y}_{\geq N+1} = \rho\, \Delta_{\geq N+1}^\mu
\,,\nn\\
\Delta_\cut &= \sqrt{1 - \rho^2}\, \Delta^\mu_{\geq N+1}
\,,\end{align}
which leads to
\begin{widetext}
\begin{equation} \label{eq:CSTrho}
\text{ST ($\rho$):} \qquad
C =
\begin{pmatrix}
(\Delta_{\geq N}^\mu)^2 + (\Delta_{\geq N+1}^\mu)^2 - 2\rho\, \Delta_{\geq N}^\mu \Delta_{\geq N+1}^\mu
&  (\rho\, \Delta_{\geq N}^\mu - \Delta^\mu_{\geq N+1}) \Delta_{\geq N+1}^\mu
 \\
(\rho\, \Delta_{\geq N}^\mu - \Delta^\mu_{\geq N+1}) \Delta_{\geq N+1}^\mu
& (\Delta^\mu_{\geq N+1})^2
\end{pmatrix}
.\end{equation}
\end{widetext}
From this one can easily see that $\rho$ corresponds to the correlation between $\Delta^\mu_{\geq N}$ and $\Delta^\mu_{\geq N+1}$. The choice $\rho = 1$ would be equivalent to the case in \eq{Cdirect}, while $\rho = 0$ reproduces \eqs{ST}{CST}. Hence, from the above arguments one should take $\rho$ to be small. In the next section, we will explore the dependence on $\rho$ in the ST method. We will see that all choices $\rho \lesssim 0.4$ give very similar results, so for our results in \sec{results} we will use the default choice $\rho = 0$.

As a final comment, note that in general one could also take $\rho$ to be a function of the binning cut. For example, at large $p^\cut$ the logarithms become small, in which case one might want to reproduce the direct scale variation uncertainties in \eq{Cdirect}. However, in this limit, typically $\Delta_{\geq N+1}^\mu$ becomes much smaller than $\Delta_{\geq N}^\mu$, which makes the precise choice of $\rho$ irrelevant there, and so it is consistent to use a fixed $\rho = 0$ everywhere.

\section{\boldmath Application to $gg\to H+2$ Jets}
\label{sec:application}

We now discuss the application of our method to the case of $pp\to H+2$ jet production via gluon fusion (which for simplicity we denote as $gg\to H+2j$, where a sum over all possible partonic channels is implied). We will study the uncertainties in the exclusive $H+2$ jet cross section as a function of two kinematic variables, $p_{THjj}$ and $\Delta\phi_{H-jj}$.

We take $\sqrt{s} = 8\TeV$ and $m_H = 125\GeV$. We use \textsc{MCFM}~\cite{Campbell:1999ah, Campbell:2006xx, Campbell:2010cz} to compute the NLO cross section, with the $ggH$ effective vertex in the infinite top mass limit. We then rescale the cross section with the exact $m_t$ dependence of the total Born cross section, $\sigma_B(m_t)/\sigma_B(\infty) = 1.0668$. We use the MSTW2008~\cite{Martin:2009bu} NLO PDFs with their corresponding value of $\alpha_s(m_Z) = 0.12018$. For all our central value predictions we use $\mu_r = \mu_f = m_H$, which was also used in Refs.~\cite{Campbell:2006xx, Campbell:2010cz}. The scale variations in the inclusive cross sections are discussed below in \subsec{inclscales}.

In our analysis we implement the 2-jet selection and VBF selection cuts summarized in Table~\ref{tab:cuts}, which are taken from the current ATLAS and CMS $H\to\gamma\gamma$ analyses. However, note that we consider the cross section for the production of an on-shell Higgs boson, without including any branching ratios or cuts on the Higgs decay products.

\begin{figure*}[t!]
\includegraphics[width=1.03\columnwidth]{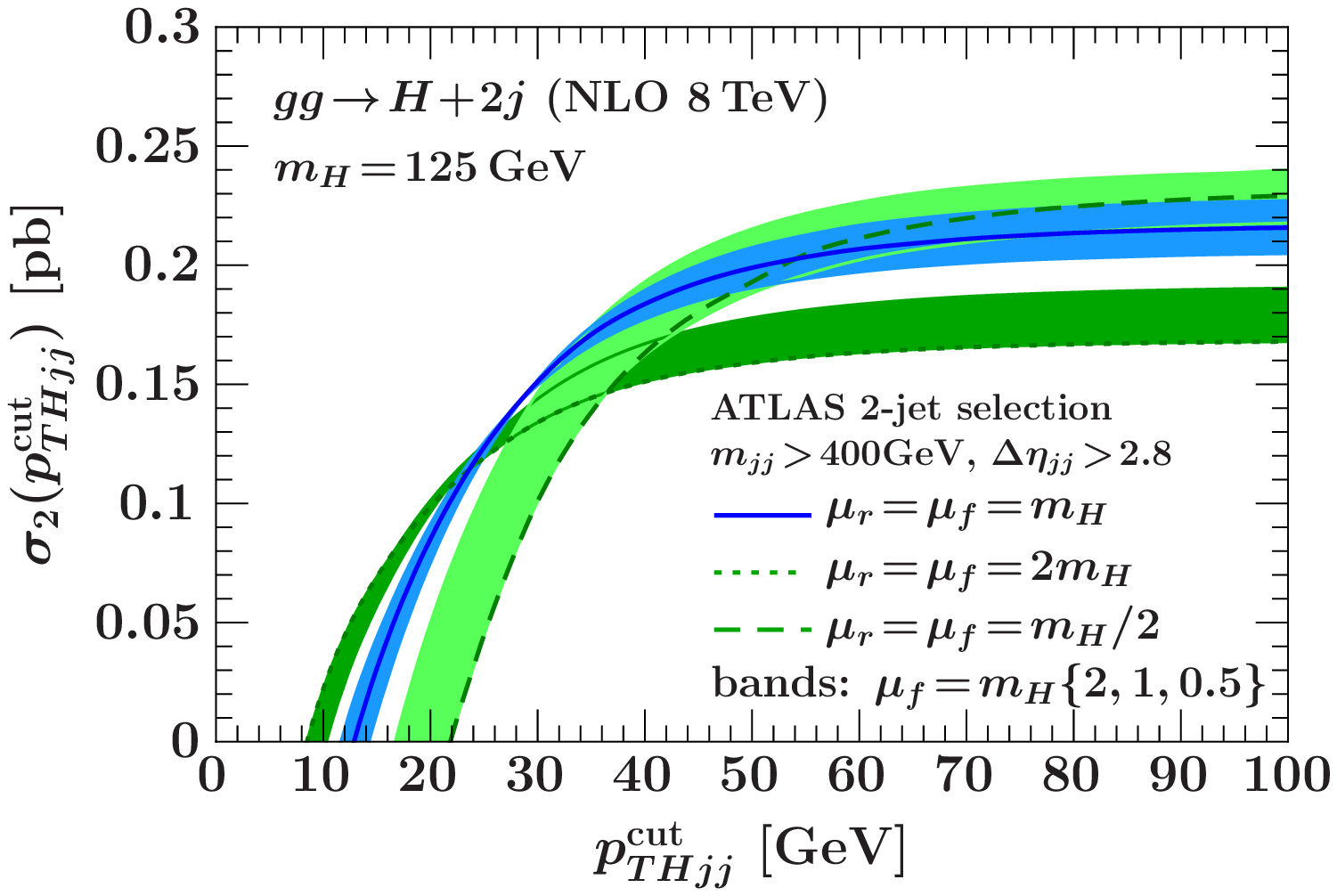}
\hfill%
\includegraphics[width=\columnwidth]{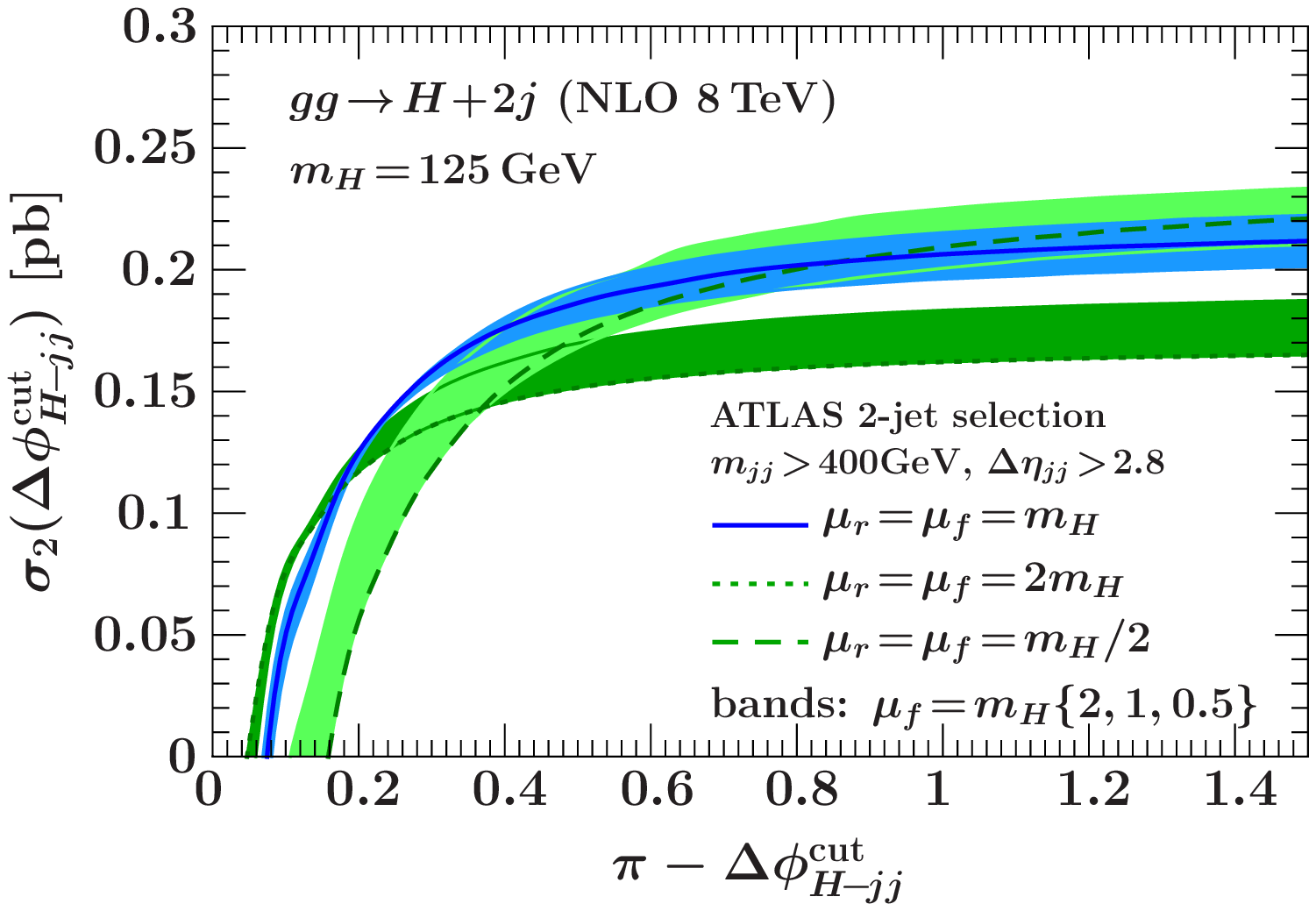}
\caption{Exclusive 2-jet cross section using the ATLAS VBF selection for various scale choices as a function of $\pT^\cut$ (left panel) and $\pi-\dphi^\cut$ (right panel).}
\label{fig:scales}
\end{figure*}

\subsection{Variables}

\subsubsection{$\pT$}

We define $p_{THjj}$ as the magnitude of the total transverse momentum of the Higgs-dijet system,
\begin{equation}
p_{THjj} = \abs{\vec p_{Tj1} + \vec p_{Tj2} + \vec p_{TH}}
\,.\end{equation}
At Born level, $\pT = 0$ and so applying a cut $\pT < \pT^\cut$ restricts the phase space to the exclusive 2-jet region.
At NLO $\pT$ is equivalent to the $p_T$ of the third jet, so it is a useful reference variable for a $p_T$-veto on additional emissions, such as the central jet vetoes applied in the $H\to WW$ and $H\to\tau\tau$ VBF analyses (see e.g. Refs.~\cite{ATLAS-CONF-2012-158, ATLAS-CONF-2012-160, CMS-PAS-HIG-12-042, CMS-PAS-HIG-12-043}).%
\footnote{The central jet veto is applied to reconstructed jets at central rapidities, which at low $p_T$ values can be heavily influenced by underlying event, pile-up, and detector effects. Since none of these effects can be accounted for by the NLO calculation, we did not attempt to study an explicit central jet veto here. Instead, we concentrate on $\pT$, which is cleaner as it only requires information about the two signal jets and the Higgs candidate.}
It is also considered directly, for example, in the latest $H\to\tau\tau$ analysis~\cite{ATLAS-CONF-2012-160}.

The exclusive 2-jet cross section $\sigma_2(\pT < \pT^\cut)$ is shown in the left panel of \fig{scales} as a function of $\pT^\cut$ and using three different combinations of the factorization and renormalization scales, $\mu_r$ and $\mu_f$. The solid line and blue band correspond to $\mu_r=m_H$ and varying $\mu_f=\{2,1,1/2\}m_H$. Similarly, we vary $\mu_f$ while keeping $\mu_r=m_H/2$ for the dark green band and $\mu_r=2m_H$ for the light green band. One can see that the biggest variation is due to the $\mu_r$ variation, while the $\mu_f$ variation only has a subdominant effect, which was already noticed in Ref.~\cite{Campbell:2006xx}. Therefore, for simplicity we will take $\mu_r = \mu_f = \mu$ and vary $\mu = \{2,1,1/2\}m_H$ when showing the direct scale variations as reference in the following.

We write the exclusive 2-jet bin defined by this cut in terms of the inclusive 2-jet cross section, $\sigma_{\geq 2}$, and the inclusive 3-jet cross section with the cut inverted as,
\begin{equation}
\sigma_2(\pT < \pT^\cut) = \sigma_{\geq 2} - \sigma_{\geq 3}(\pT > \pT^\cut)
\,,\end{equation}
where in all cases the remaining VBF selection cuts in Table~\ref{tab:cuts} are applied (excluding the cut on $\dphi$ in this case).

The restriction on $\pT$ is infrared sensitive and induces Sudakov logarithms of the form $L =\ln(\pT^\cut/m_H)$ in the perturbative series of $\sigma_2$ and $\sigma_{\geq 3}$. In \fig{scales} we see that the veto starts to have a noticeable effect below $\pT \lesssim 50\GeV$, where the different scale variations start crossing and we start to see cancellations between $\sigma_{\geq 2}$ and $\sigma_{\geq 3}$.
In the region below $p_{THjj} \lesssim 20\GeV$, the logarithms have grown large enough for the NLO cross section to go negative and the fixed-order perturbative expansion to break down. In the intermediate region in between, the fixed-order prediction can still be used, but the direct scale variation does not provide a reliable uncertainty estimate as it does not properly take into account the effect of the binning cut.

\subsubsection{$\dphi$}

As shown in Table~\ref{tab:cuts}, the VBF category in the $H \rightarrow \gamma\gamma$ analyses by ATLAS and CMS includes a cut $\Delta\phi_{H-jj} > 2.6$ radians $(150\deg)$ (where the Higgs momentum is represented by the total momentum of the diphoton system). Taking the beam direction along the $z$-axes, $\Delta\phi_{H-jj}$ is defined as
\begin{equation}
\cos \dphi = \frac{(\vec p_{Tj1} + \vec p_{Tj2}) \cdot \vec p_{TH}}{\abs{\vec p_{Tj1}+ \vec p_{Tj2}} \abs{\vec p_{TH}}}
\,.\end{equation}
Momentum conservation in the transverse plane implies that events with only two jets always have $\dphi \approx \pi$, so the constraint $\dphi > \dphi^\cut$ forces the kinematics into the exclusive 2-jet region and restricts additional emissions. Hence, it behaves similar to $\pT^\cut$ and for $\pi - \dphi^\cut\!\to 0$ induces large logarithms in the perturbative series. The exclusive 2-jet cross section in terms of $\dphi^\cut$ is written as
\begin{equation}
\sigma_2(\dphi > \dphi^\cut) = \sigma_{\geq 2} - \sigma_{\geq 3}(\dphi < \dphi^\cut)
\,,\end{equation}
with the remaining VBF cuts applied in all three cross sections. The right panel of \fig{scales} shows $\sigma_2(\dphi > \dphi^\cut)$ plotted as a function of $\pi - \dphi^\cut$, where one can clearly see the very similar behavior to the $\pT^\cut$ case in the left panel. Here, the exclusive cut on $\dphi$ starts having a noticeable effect below $\pi - \dphi \lesssim 0.6$, and the fixed-order perturbative expansion breaks down below around $\pi-\Delta\phi \lesssim 0.2$. In the transition region in between, the direct scale variations again do not provide a meaningful uncertainty estimate, because they neglect the effect of $\Delta_\cut$.

\subsection{Inclusive Scale Uncertainties}
\label{subsec:inclscales}

\begin{figure*}
\includegraphics[width=1.03\columnwidth]{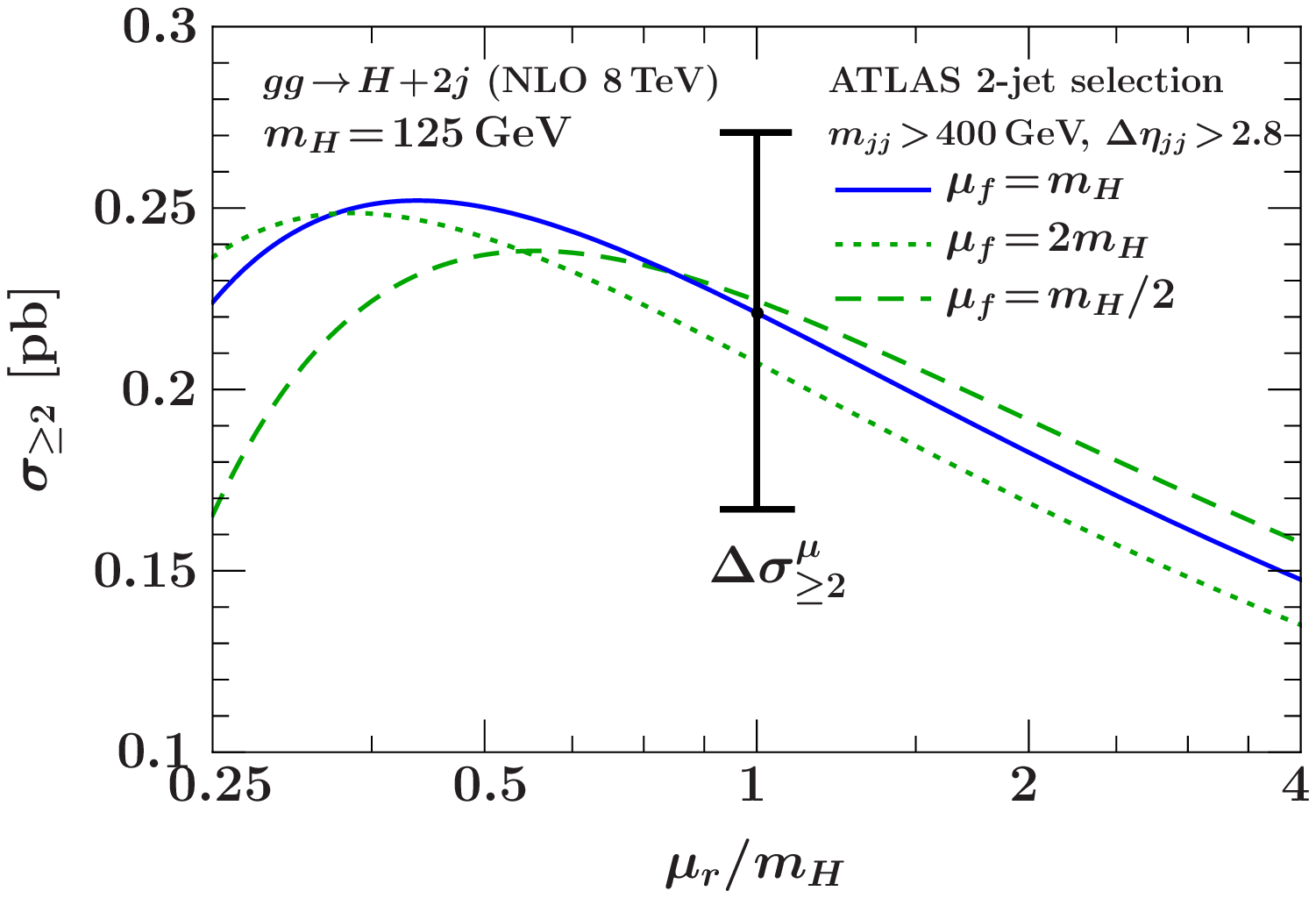}%
\hfill%
\includegraphics[width=\columnwidth]{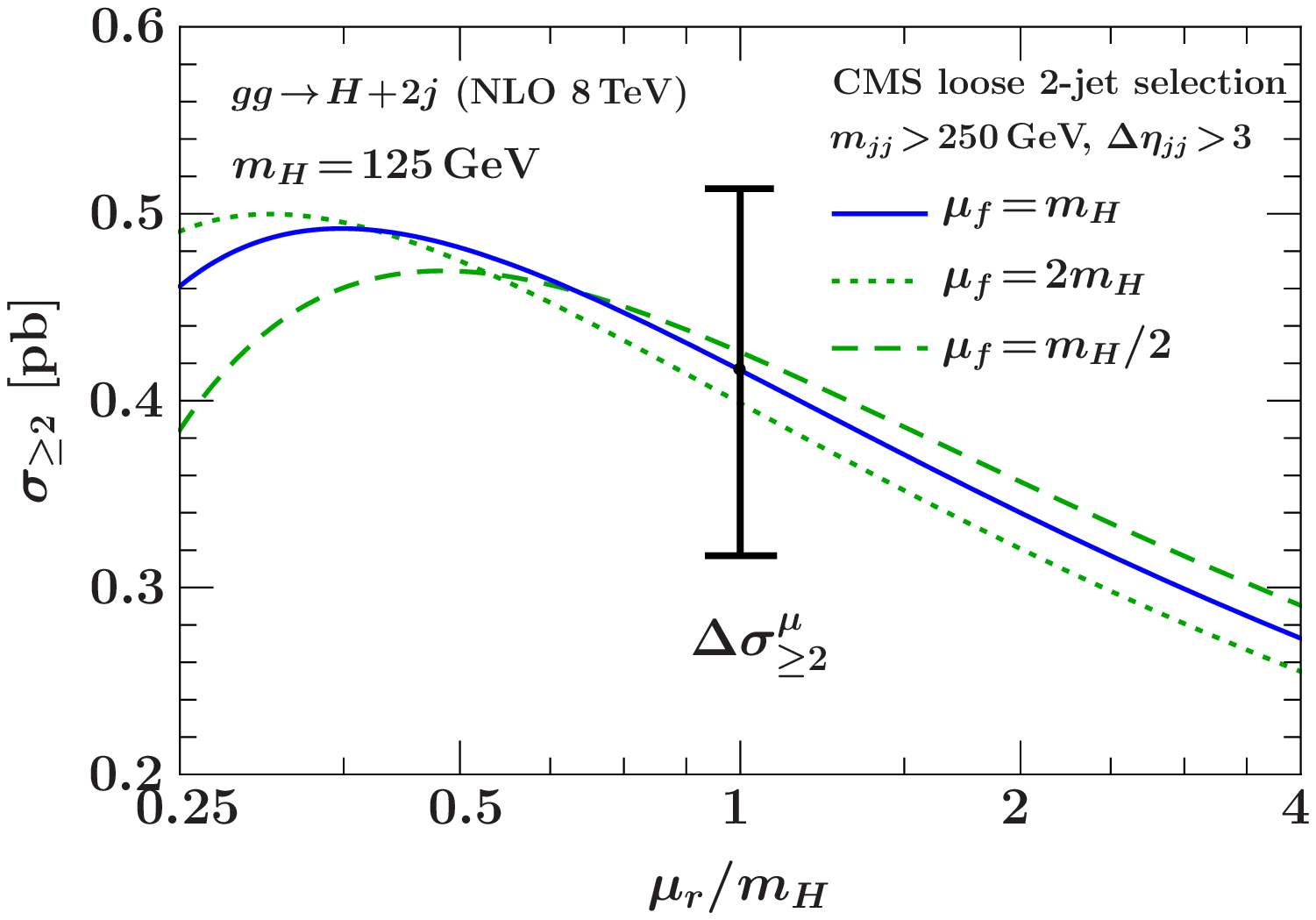}%
\caption{Inclusive 2-jet cross section over a range of $\mu_r/m_H$ for ATLAS VBF selection (left panel) and  CMS loose selection (right panel). The three curves show different values of $\mu_f$. The blue solid, green dotted, and green dashed curves correspond to $\mu_f=m_H$, $\mu_f=2m_H$, and $\mu_f=m_H/2$, respectively. The uncertainty bars show the inclusive 2-jet scale variation uncertainty.}
\label{fig:murmuf2}
\end{figure*}

\begin{figure*}
\includegraphics[width=1.03\columnwidth]{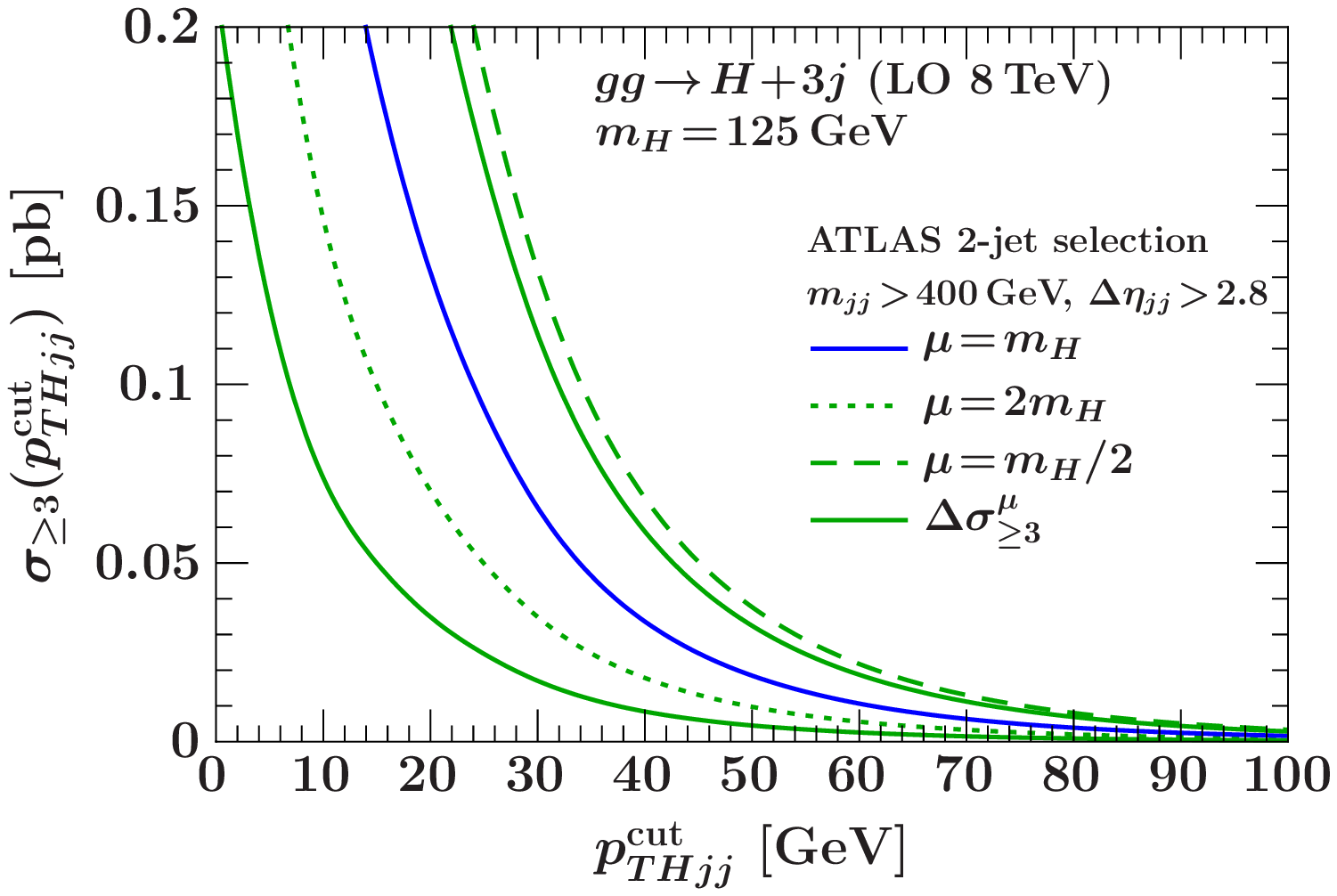}%
\hfill%
\includegraphics[width=\columnwidth]{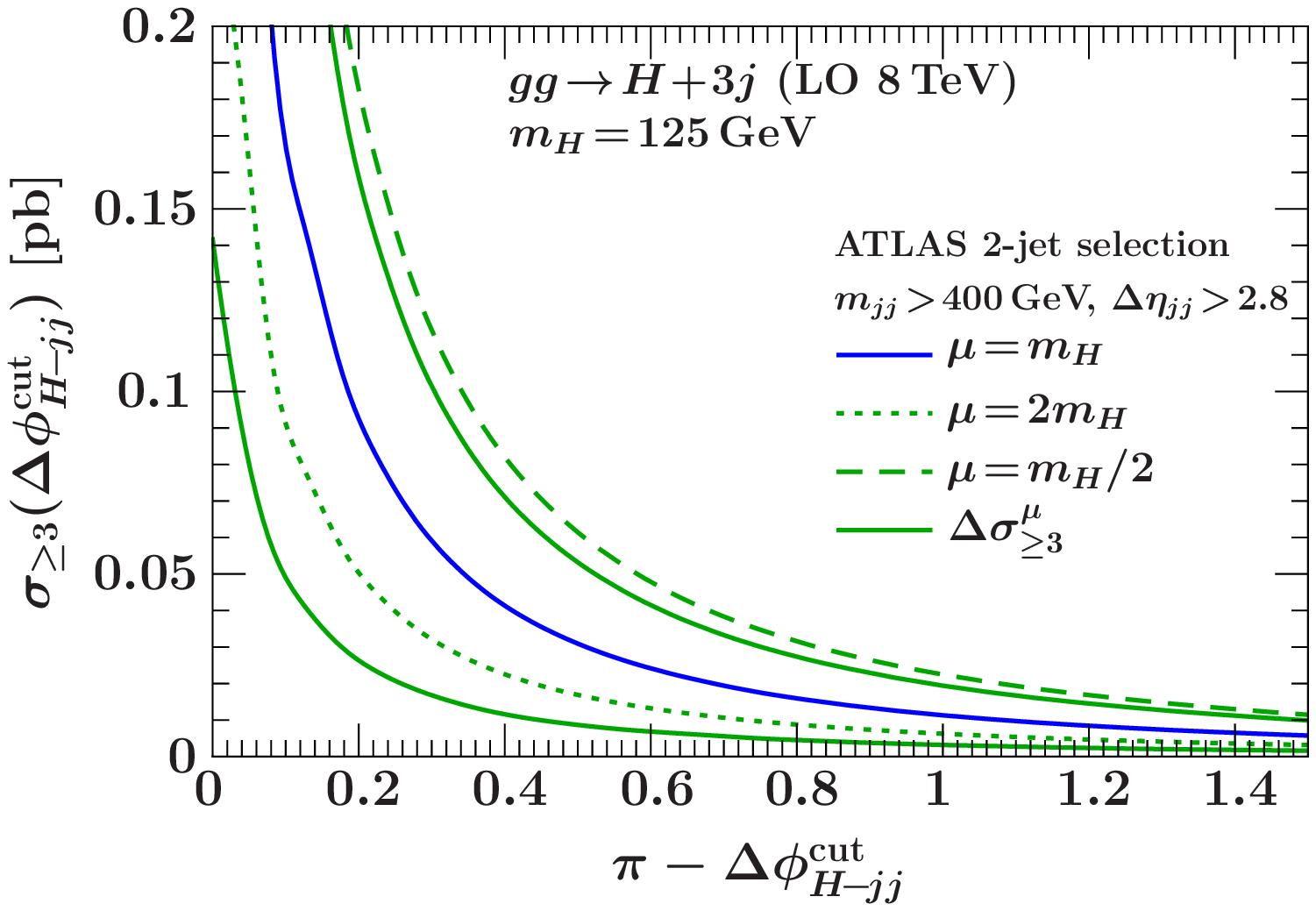}%
\caption{Inclusive 3-jet cross section as a function of $\pT^\cut$ (left panel) and $\pi -\dphi^\cut$ (right panel) for the ATLAS VBF selection.
The outer solid green lines show the inclusive 3-jet scale variation uncertainty after symmetrization.}
\label{fig:murmuf3}
\end{figure*}

The two fixed-order scale variation uncertainties we require as inputs are $\Delta_{\geq 2}^\mu$ and $\Delta_{\geq 3}^\mu$. In \fig{scales}, one can already see that the scale variation is asymmetric at large values of $\pT^\cut$ and $\pi-\dphi^\cut$. In \fig{murmuf2}, we show the scale dependence of the inclusive 2-jet cross section, $\sigma_{\geq 2}$, where we plot it over a range of $1/4<\mu_r/m_H<4$ for three different values of $\mu_f$. We take  $\mu_f=\mu_r=m_H$, corresponding to the $\mu_r/m_H=1$ point on the blue solid line, as our central value for $\sigma_{\geq 2}$, and consider the range $0.5 \leq \mu_r/m_H \leq 2$ to estimate the inclusive scale uncertainty. The maximum deviation from the central value is given by the green dotted curve for $\mu_f = \mu_r = 2m_H$. We use this maximum variation to construct a symmetric uncertainty $\Delta_{\geq 2}^\mu$, as shown by the uncertainty bar in the figure. It corresponds to a relative uncertainty at NLO of $21\%$, which is similar to what was found in earlier studies~\cite{Campbell:2006xx, Campbell:2010cz} where a somewhat looser VBF selection was used. The corresponding uncertainty at LO is $+76\%$ and $-40\%$.

In \fig{murmuf3}, we illustrate the scale variation uncertainties for the inclusive 3-jet cross section, $\sigma_{\geq 3}$, for both $\pT^\cut$ and $\dphi^\cut$, and using the ATLAS selection as example. (The results for $\sigma_{\geq 3}$ with the CMS selections look very similar except for the different overall scale.) The blue solid line shows the cross section for $\mu_r=\mu_f=m_H$, which we take as the central value for $\sigma_{\geq 3}$. The green dashed and dotted lines show the scale variations $\mu_r=\mu_f=m_H/2$ and $\mu_r=\mu_f=2m_H$, respectively. For simplicity, we symmetrize the uncertainty by taking half of the difference between the up and down variations as the inclusive 3-jet scale uncertainty $\Delta_{\geq 3}^\mu$, i.e., we keep the size of the band and move it to be symmetric about the central blue line, which is shown by the outer solid green lines. The relative uncertainty is of $\ord{70\%}$ and almost independent of $\pT^\cut$ and $\dphi^\cut$. This rather large uncertainty is not too surprising, since this is a leading-order $H+3j$ cross section, which starts at $\alpha_s^5$.

\subsection{Exclusive Uncertainty}

Having obtained the perturbative uncertainties $\Delta_{\geq 2}^\mu$ and $\Delta_{\geq 3}^\mu$ in the inclusive cross sections from the usual scale variation, we now study the resulting uncertainty $\Delta_2$ in the exclusive 2-jet cross section according to the discussion in \sec{jetbinreview}. From \eq{CSTrho} we have in general
\begin{equation} \label{eq:13}
\Delta_2^2 = (\Delta^\mu_{\geq 2})^2 + (\Delta^\mu_{\geq 3})^2 - 2\rho\, \Delta^\mu_{\geq 2} \Delta^\mu_{\geq 3}
\,,\end{equation}
where $\rho$ is the assumed correlation between $\Delta_{\ge2}^\mu$ and $\Delta_{\ge3}^\mu$. 

\begin{figure*}[t!]
\includegraphics[width=1.03\columnwidth]{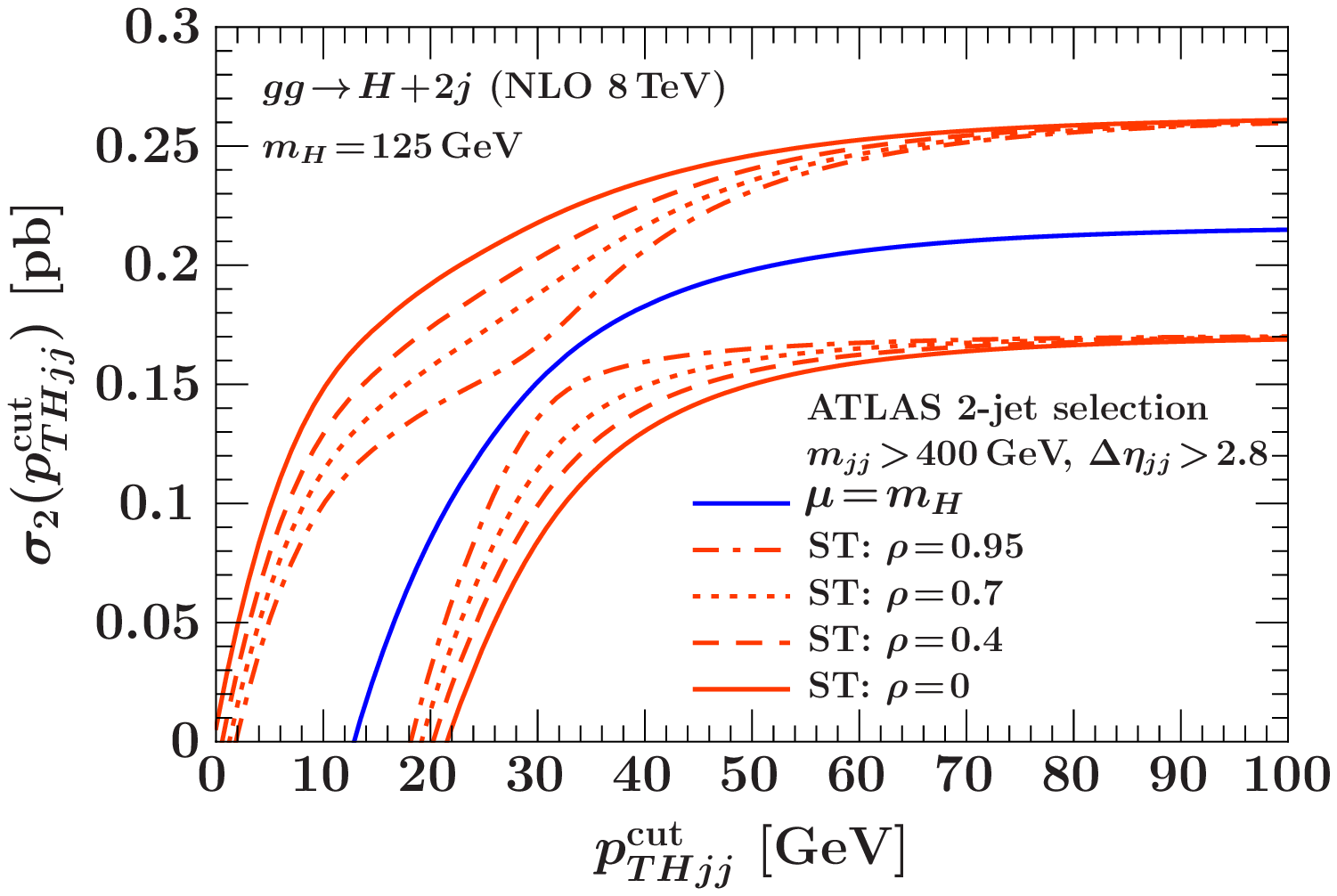}%
\hfill%
\includegraphics[width=\columnwidth]{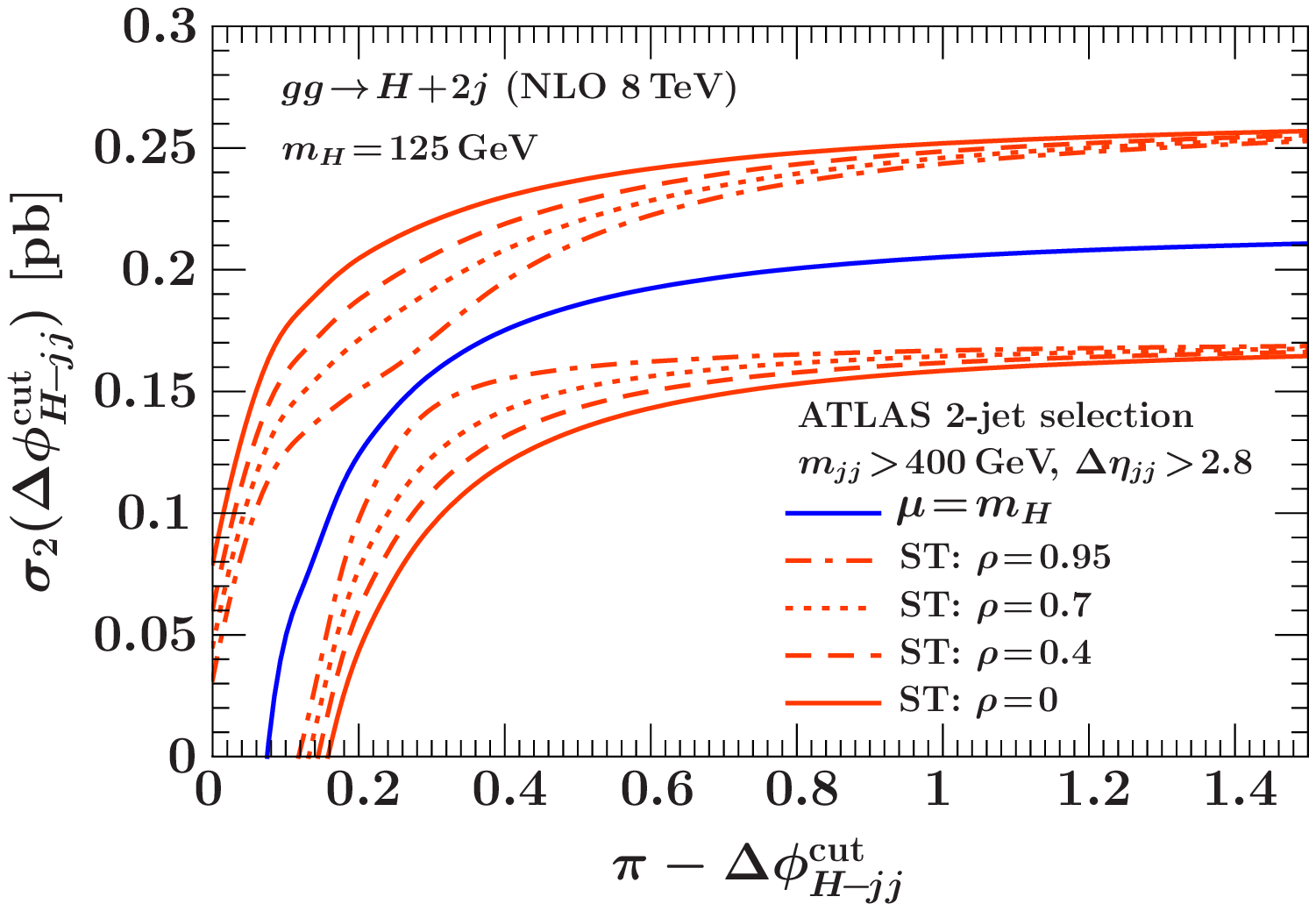}%
\caption{Perturbative uncertainties in the exclusive 2-jet cross section with the ATLAS VBF selection as a function of $\pT^\cut$ (left panel) and $\pi-\dphi^\cut$ (right panel) for different choices of the correlation parameter $\rho$. Our default choice is $\rho = 0$.}
\label{fig:rho}
\end{figure*}

\subsubsection{Dependence on $\rho$}

We first investigate the dependence on the choice of $\rho$. In \fig{rho} we show the uncertainty in the exclusive 2-jet cross section as a function of $\pT^\cut$ and $\dphi^\cut$ for different values of $\rho$ from $0$ to $0.95$. The outermost solid curves show the uncertainty obtained with our default choice $\rho=0$, which effectively assumes that $\Delta^\mu_{\ge 2}$ and $\Delta^\mu_{\ge3}$ are uncorrelated. For $\rho \lesssim 0.4$ the results are not very sensitive to the precise value of $\rho$, which is reassuring and shows that $\rho=0$ is in fact a safe choice on the conservative side.

As $\rho$ increases further, the uncertainty bands in the transition region keep shrinking, and for $\rho=0.95$, shown by the innermost dot-dashed lines, pinch near $\pT^\cut \simeq 30\GeV$ and $\pi-\dphi^\cut \simeq 0.3$. (For $\rho = 1$ the uncertainty goes exactly to zero around these points.) This is because for $\rho \to 1$, $\Delta^\mu_{\ge3}$ and $\Delta^\mu_{\ge 2}$ become $100\%$ correlated, which is equivalent to the case of direct scale variation. (The only difference compared to the direct scale variations we saw in \fig{scales} is that here we symmetrized the scale variations.)

One can also see that for large cut values, where the veto is not relevant and we approach the inclusive $2$-jet cross section, the choice of $\rho$ becomes irrelevant, because the absolute size of $\Delta_{\geq 3}^\mu$ becomes numerically negligible compared to $\Delta_{\geq 2}^\mu$.

\subsubsection{Comparison to Efficiency Method}

\begin{figure*}[t!]
\includegraphics[width=1.03\columnwidth]{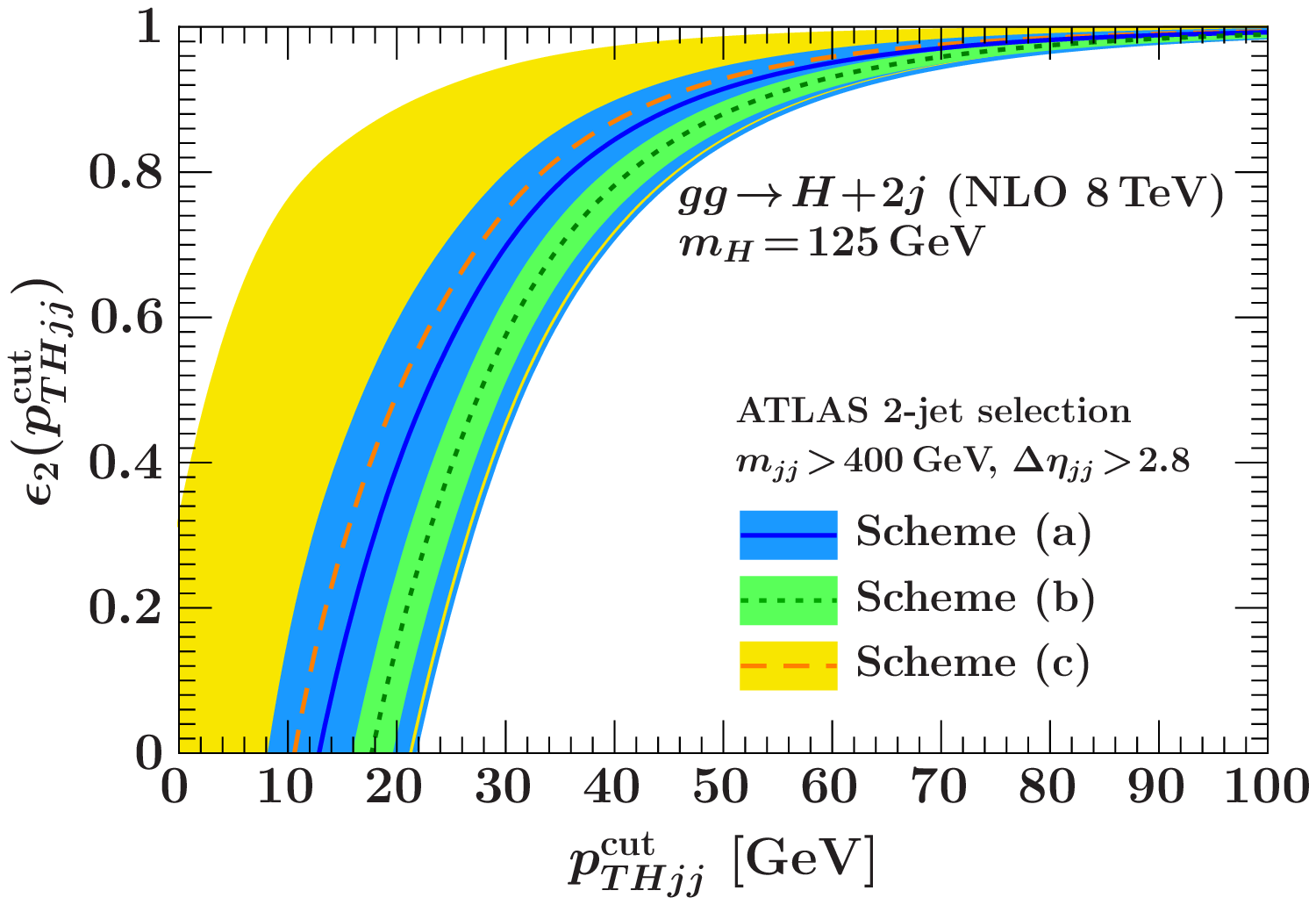}%
\hfill%
\includegraphics[width=\columnwidth]{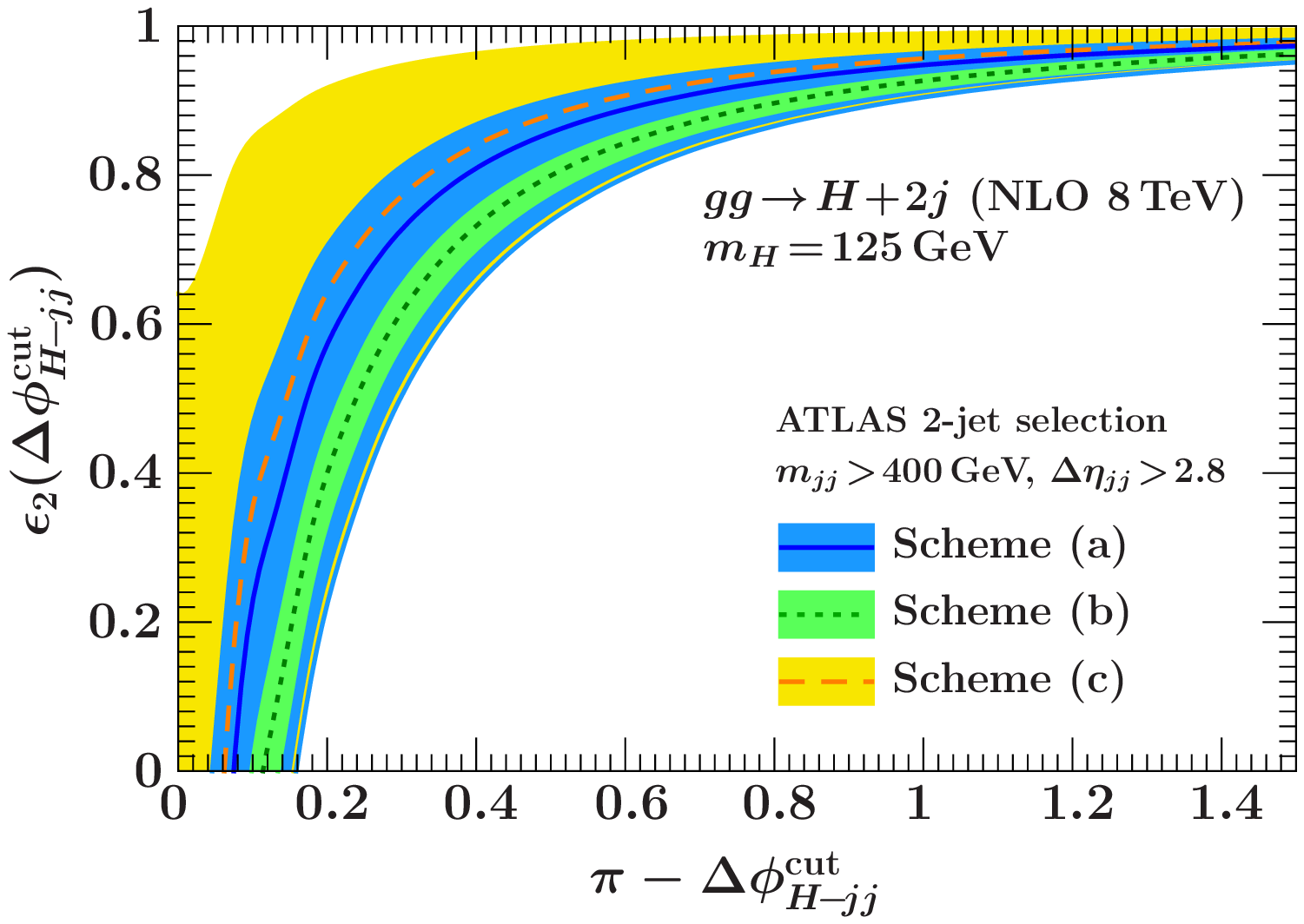}%
\caption{Exclusive 2-jet efficiency for different schemes in the efficiency method for $\pT^\cut$ (left panel) and $\pi-\dphi^\cut$ (right panel) using the ATLAS VBF selection.}
\label{fig:efficiency}
\end{figure*}

\begin{figure*}[t!]
\includegraphics[width=1.03\columnwidth]{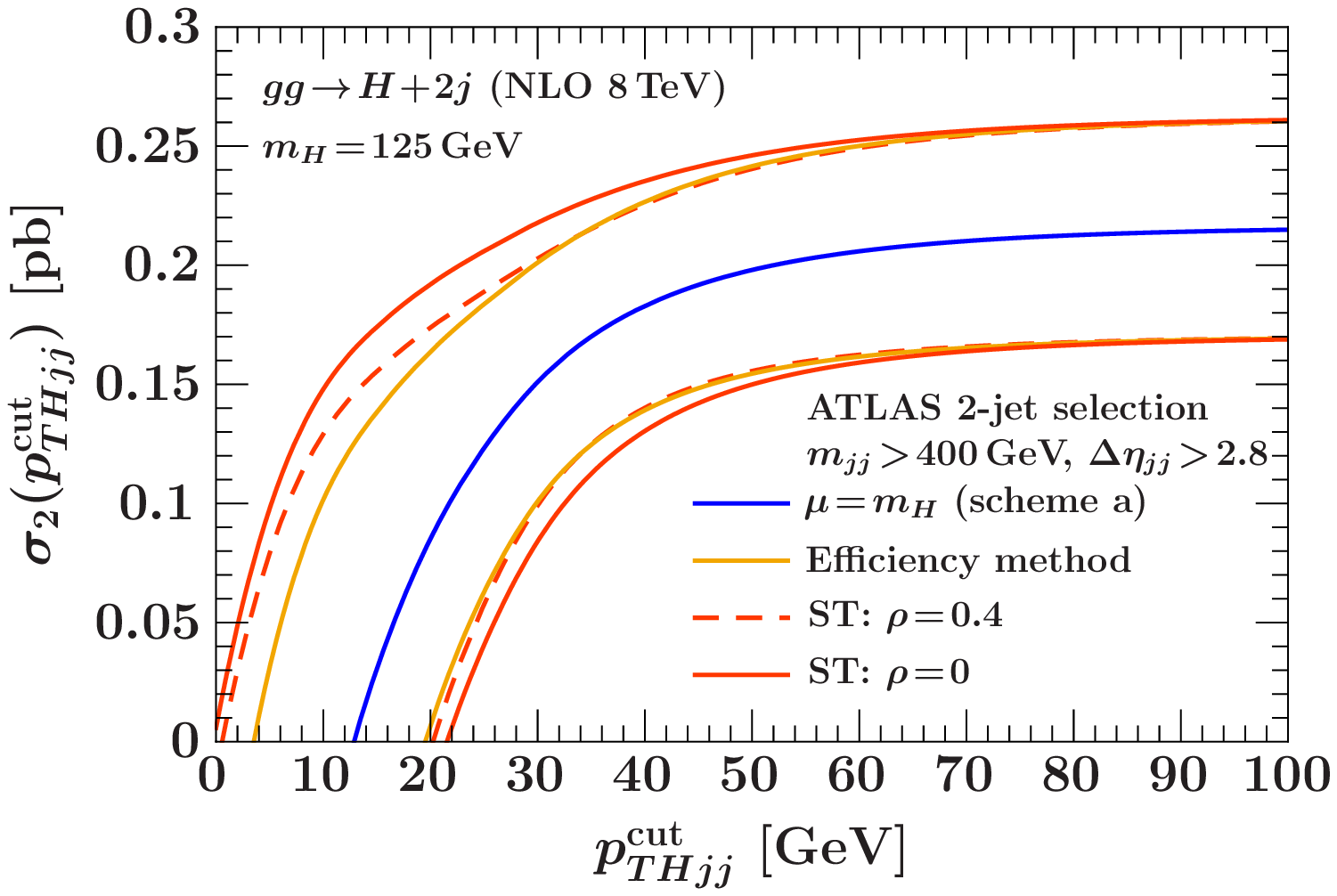}%
\hfill%
\includegraphics[width=\columnwidth]{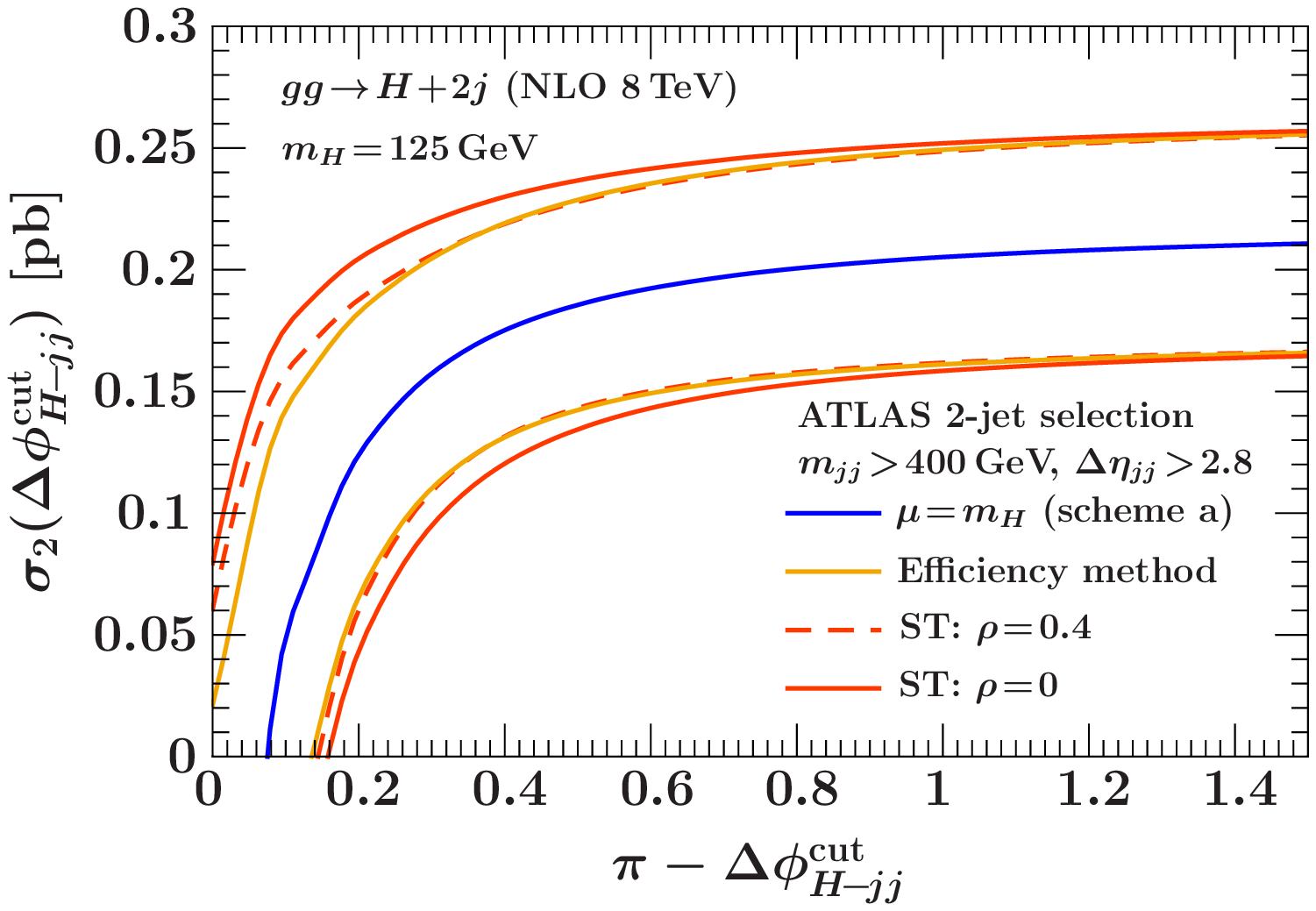}%
\caption{Comparison of the ST method with the efficiency method for $\pT^\cut$ (left panel) and $\pi-\dphi^\cut$ (right panel) using the ATLAS VBF selection. The exclusive scale uncertainties from both methods are consistent with each other. The uncertainties from the efficiency method are very close to those from the ST method with $\rho = 0.4$.}
\label{fig:effST}
\end{figure*}

Another prescription to obtain fixed-order uncertainty estimates for exclusive jet cross section, which is based on using veto efficiencies, was applied in Ref.~\cite{Banfi:2012yh} to the $0$-jet case at NNLO. We will refer to it as ``efficiency method''.

In Ref.~\cite{AlcarazMaestre:2012vp} it was shown that for the case of $H+0$ jets at NNLO the ST method and efficiency method yield very similar uncertainties, providing a good cross check on both methods.

The starting point in the efficiency method is to write the exclusive jet cross section in terms of the corresponding inclusive jet cross section times the corresponding exclusive efficiency, i.e., applied to our 2-jet case,
\begin{align}
\sigma_2 &= \sigma_{\geq 2} \Bigl(1 - \frac{\sigma_{\geq 3}}{\sigma_{\geq 2}} \Bigr)
\equiv \sigma_{\geq 2} \times \epsilon_2
\,,\nn\\
\sigma_{\geq 3} &= \sigma_{\geq 2}\, (1 - \epsilon_2)
\,,\end{align}
where the logarithmic series induced by the jet binning now only affects the efficiency. The basic assumption~\cite{Banfi:2012yh} one then makes is to treat the perturbative uncertainties in $\sigma_{\geq 2}$ and $\epsilon_2$ as uncorrelated (which one can think of as a multiplicative version of the ST approach). One should be aware that this method does not satisfy one of our starting conditions, namely the total uncertainty $\Delta_{\geq 3}$ for $\sigma_{\geq3}$ will not be given by its standard scale variation $\Delta_{\geq 3}^\mu$ anymore. Nevertheless, it is a useful way to gain additional insights into the size of higher-order corrections.

The 2-jet efficiency $\epsilon_2 = 1 - \sigma_{\geq 3}/\sigma_{\geq 2}$ is still an exclusive quantity. Similar cancellations between the two perturbative series for $\sigma_{\geq 2}$ and $\sigma_{\geq 3}$ can happen in their ratio than in their difference, so the direct scale variation for $\epsilon_2$ might not provide a reliable uncertainty estimate. To circumvent this, in Ref.~\cite{Banfi:2012yh} the perturbative uncertainty in $\epsilon$ is instead estimated by using three different schemes for how to write the perturbative result for $\epsilon$, which are all equivalent up to the desired order in $\alpha_s$, but differ in the higher-order terms that are retained or not.

The inclusive 2-jet and 3-jet cross sections have the following perturbative structure
\begin{align}
\sigma_{\ge2} &= \alpha_s^2 \bigl[\sigma_{\ge2}^{(0)} + \alpha_s\, \sigma_{\ge2}^{(1)} + \alpha_s^2\, \sigma_{\ge2}^{(2)} + \ord{\alpha_s^3} \bigr]
\,,\nn\\
\sigma_{\ge3} &= \alpha_s^2 \bigl[\alpha_s \sigma_{\ge3}^{(0)} + \alpha_s^2\, \sigma_{\ge3}^{(1)}  + \ord{\alpha_s^3} \bigr]
\,.\end{align}
At NLO, the pieces we have available are $\sigma_{\ge2}^{(0)}$, $\sigma_{\ge2}^{(1)}$, and $\sigma_{\ge3}^{(0)}$. In scheme (a) one defines the efficiency by keeping the full expressions in numerator and denominator, which at NLO gives
\begin{equation}
\epsilon_2^{(a)}
= 1 - \frac{\sigma_{\ge3}}{\sigma_{\ge 2}}
= 1 - \frac{\alpha_s \sigma^{(0)}_{\ge 3}}{\sigma^{(0)}_{\ge 2} + \alpha_s\sigma^{(1)}_{\ge2}} + \ord{\alpha_s^2}
\,.\end{equation}
In scheme (b) one keeps the same number of terms in the perturbative series in the denominator as in the numerator, which in our case amounts to dropping the $\sigma_{\geq 2}^{(1)}$ term in the denominator,
\begin{equation}
\epsilon_2^{(b)} = 1- \alpha_s\,\frac{\sigma^{(0)}_{\ge 3}}{\sigma^{(0)}_{\ge 2}} + \ord{\alpha_s^2}
\,.\end{equation}
Finally, in scheme (c) one strictly reexpands the ratio to a given order in $\alpha_s$, which to $\ord{\alpha_s}$ unfortunately yields the same result as scheme (b). To produce another expression with differing higher-order terms, the closest scheme (c) analog we can do is to keep the $\ord{\alpha_s^2}$ cross term that comes from expanding the denominator, so
\begin{equation}
\epsilon_2^{(c)}
= 1 - \alpha_s\, \frac{\sigma^{(0)}_{\ge 3}}{\sigma^{(0)}_{\ge 2}}
\biggl(1 - \alpha_s\, \frac{\sigma_{\geq 2}^{(1)}}{\sigma^{(0)}_{\geq 2}} \biggr) + \ord{\alpha_s^2}
\,.\end{equation}

In \fig{efficiency} we show the result for $\epsilon_2$ in the three schemes for both $\pT^\cut$ and $\dphi^\cut$ using the ATLAS VBF selection. The central lines show the results for $\mu_r = \mu_f = \mu = m_H$, while the bands are obtained from varying $\mu = \{2, 1/2\}m_H$ in each scheme. At NLO the central values from the three schemes are quite close and still lie within the direct scale variation of scheme (a), so their difference does not provide a useful uncertainty estimate here. The direct scale variation in scheme (b) is very small and in scheme (c) abnormally large (which is very similar to what was seen in Ref.~\cite{Banfi:2012yh}). Hence, in the end the most reasonable choice to get an uncertainty estimate for $\epsilon_2$ is to just use the direct scale variation in scheme (a).

In \fig{effST} we compare the results of the ST and efficiency methods for the exclusive 2-jet cross section $\sigma_2$ for both $\pT^\cut$ and $\dphi^\cut$ using the ATLAS VBF selection. The blue solid curve shows our usual NLO central value, which is equivalent to the central value from scheme (a). The light orange solid curves are the uncertainties obtained in the efficiency method by combining the scale uncertainties $\Delta_{\geq2}^\mu$ with the direct scale variations in $\epsilon_2^{(a)}$ treating both as uncorrelated. The dark orange solid curves show the default ST uncertainties for $\rho = 0$, which are somewhat larger. The dashed lines show the ST uncertainties for $\rho = 0.4$, which agree almost perfectly with the efficiency method. This result is not surprising. Basically, to obtain the $\epsilon_2$ scale uncertainty we vary the scales correlated in $\sigma_{\geq2}$ and $\sigma_{\geq 3}$, which has the effect of reintroducing a certain amount of correlation between $\Delta_{\geq 2}^\mu$ and $\Delta_{\geq 3}^\mu$ when considering $\sigma_2$, which is also what a nonzero value of $\rho$ does. Overall, the good consistency between the various methods gives us confidence in the reliability of our uncertainty estimates.

\begin{figure*}[t!]
\subfigure[ATLAS VBF selection]
{\includegraphics[width=1.03\columnwidth]{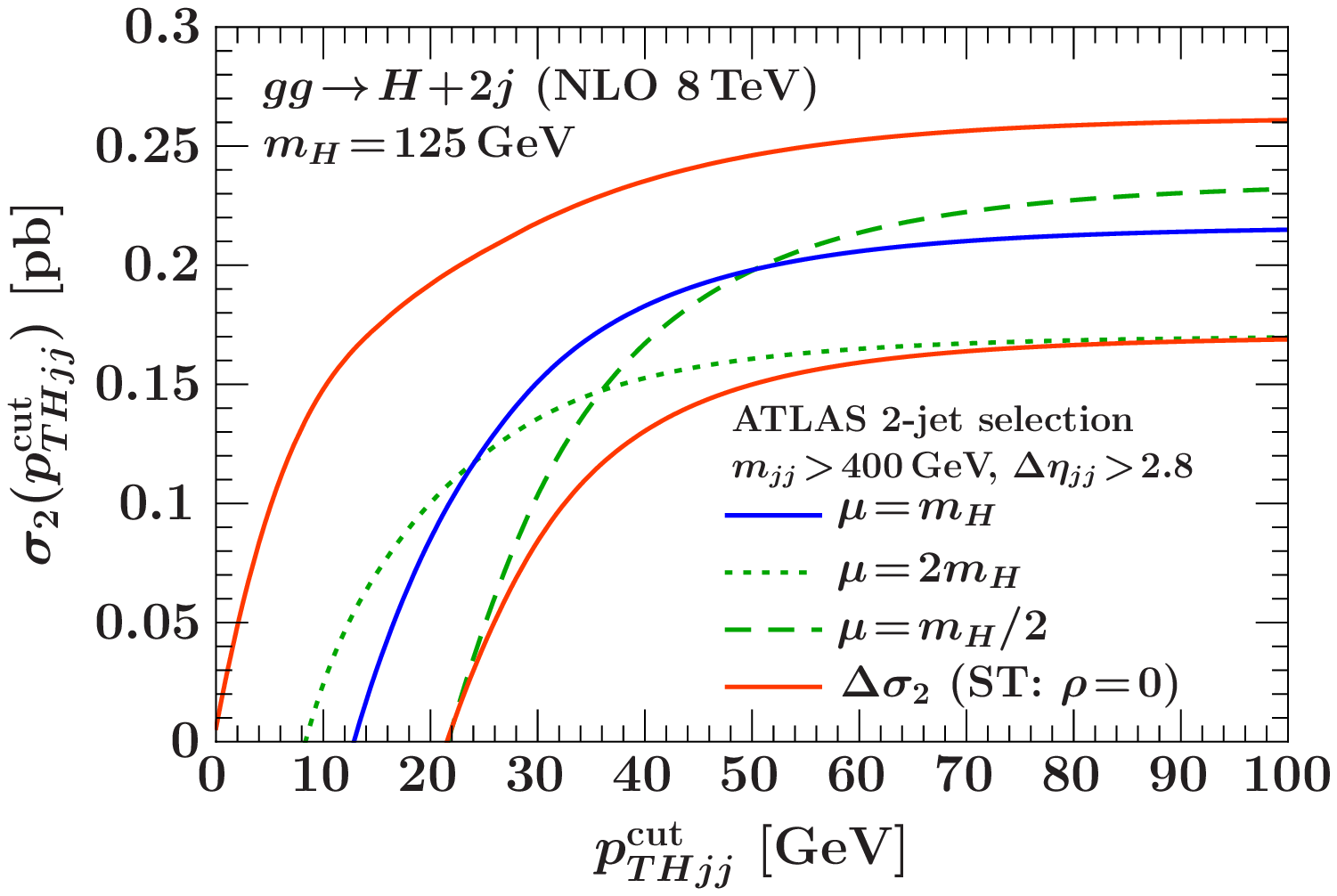}\label{fig:atlaspt}}%
\hfill%
\subfigure[ATLAS VBF selection]
{\includegraphics[width=\columnwidth]{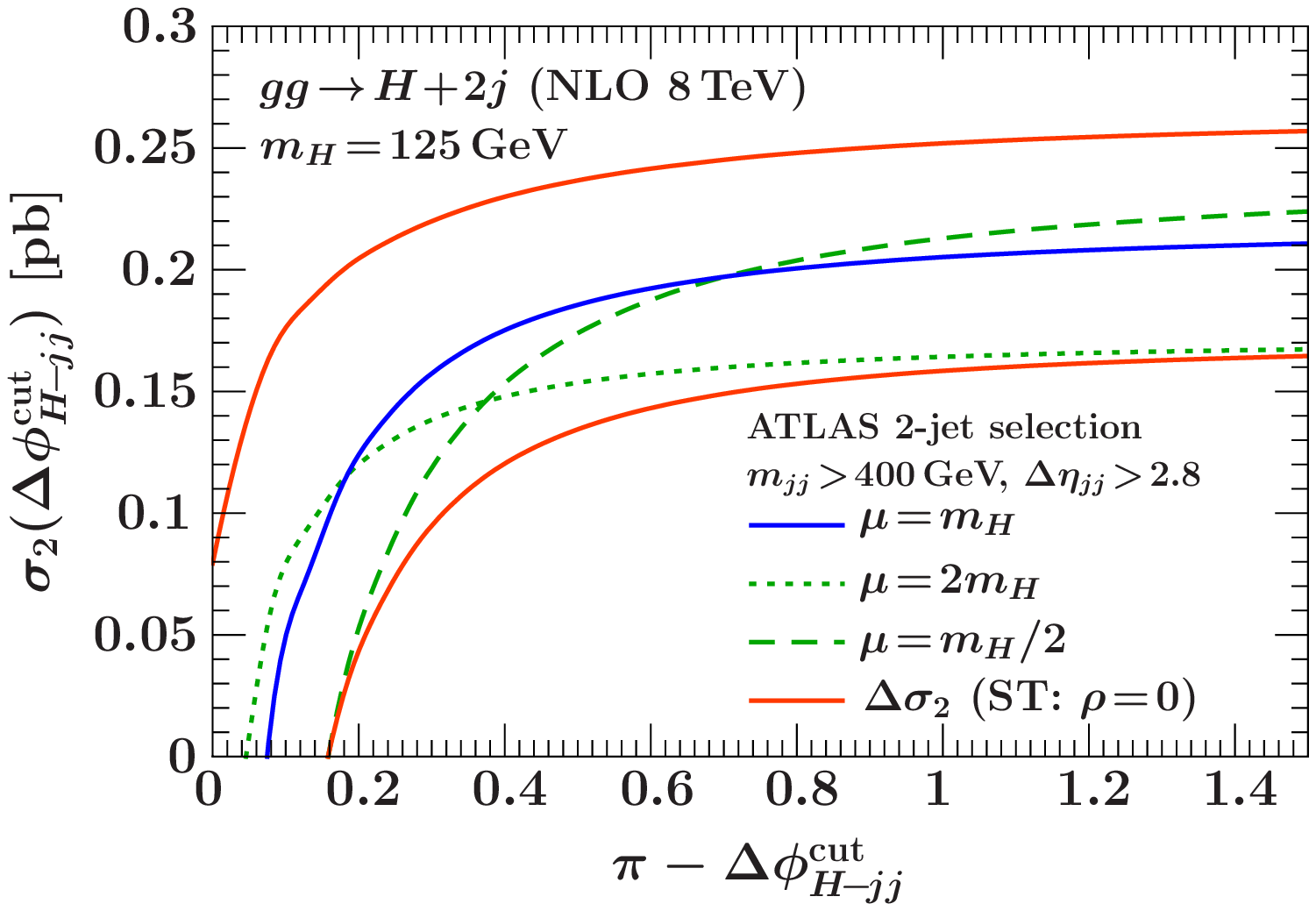}\label{fig:atlasphi}}%
\\
\subfigure[CMS loose VBF selection]
{\includegraphics[width=1.03\columnwidth]{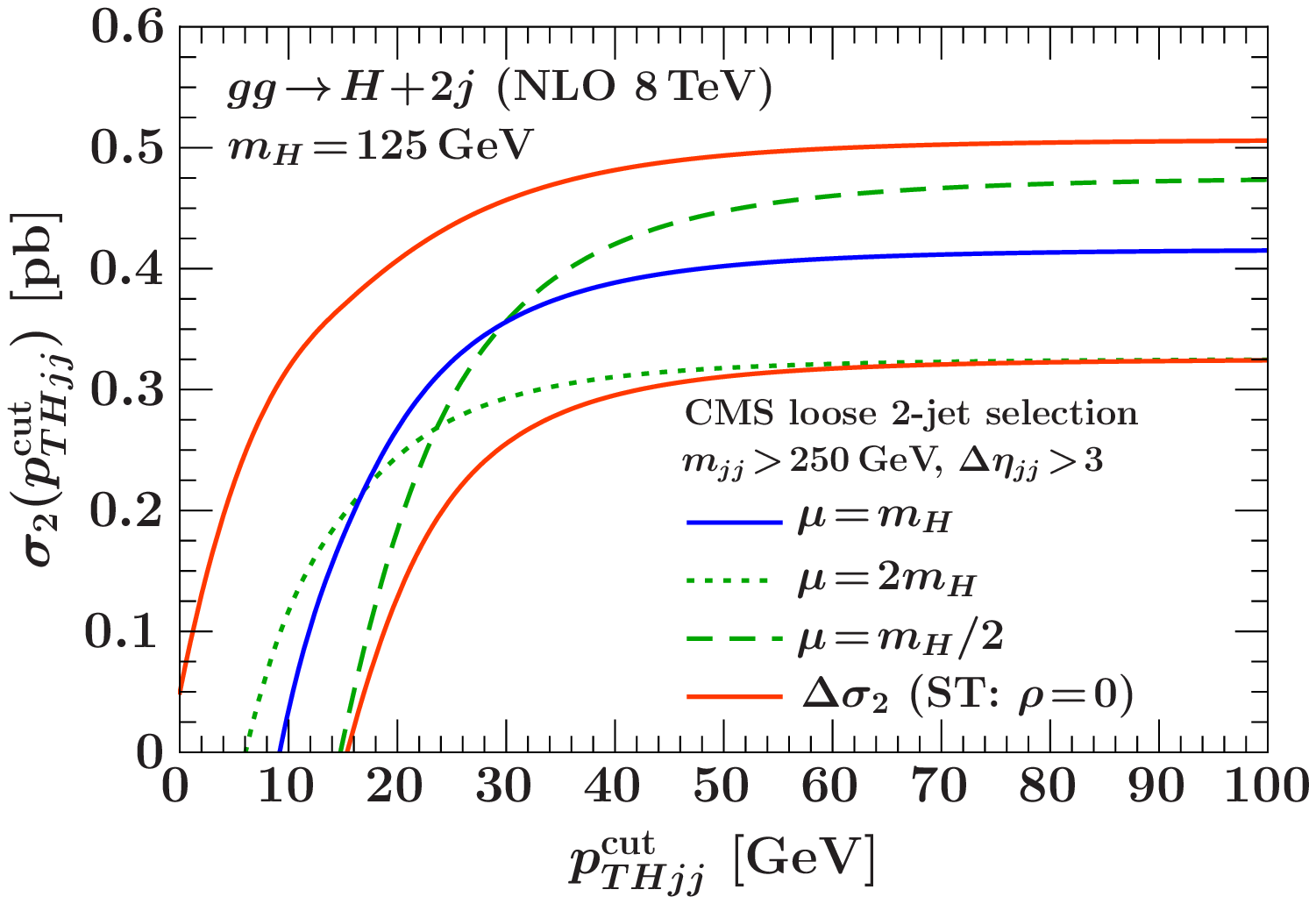}\label{fig:cmsloosept}}%
\hfill%
\subfigure[CMS loose VBF selection]
{\includegraphics[width=\columnwidth]{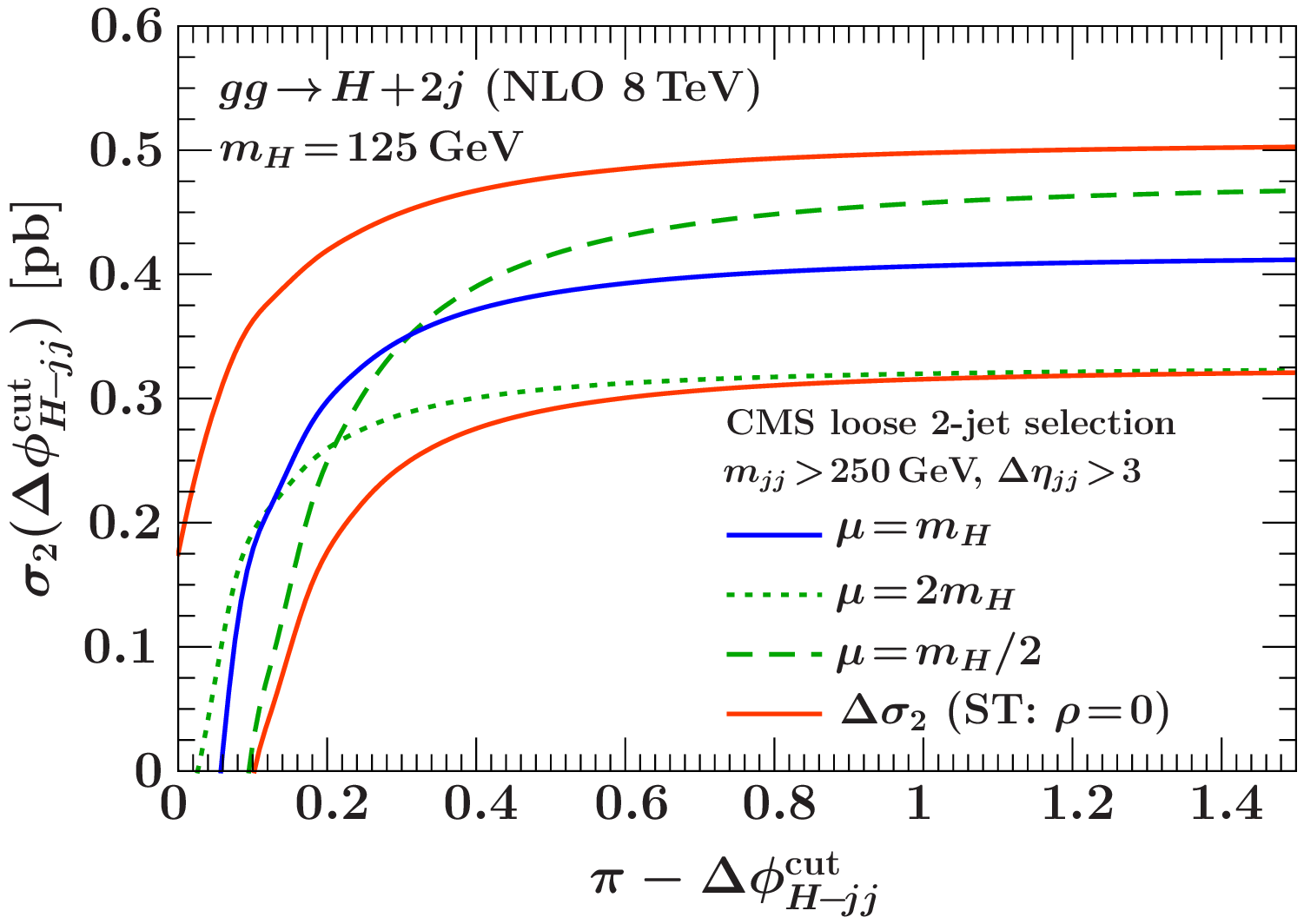}\label{fig:cmsloosephi}}%
\\
\subfigure[CMS tight VBF selection]
{\includegraphics[width=1.03\columnwidth]{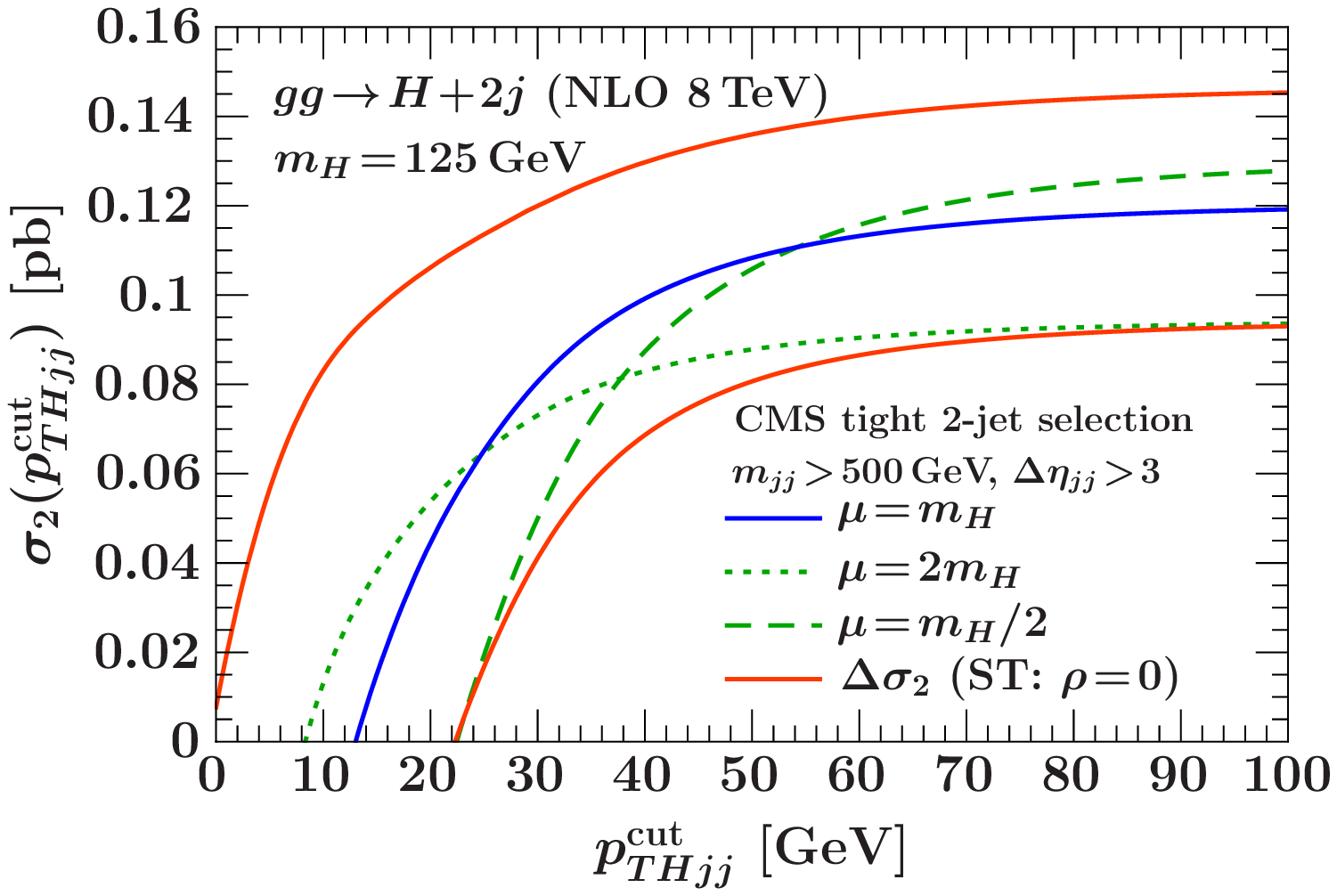}\label{fig:cmstightpt}}%
\hfill%
\subfigure[CMS tight VBF selection]
{\includegraphics[width=\columnwidth]{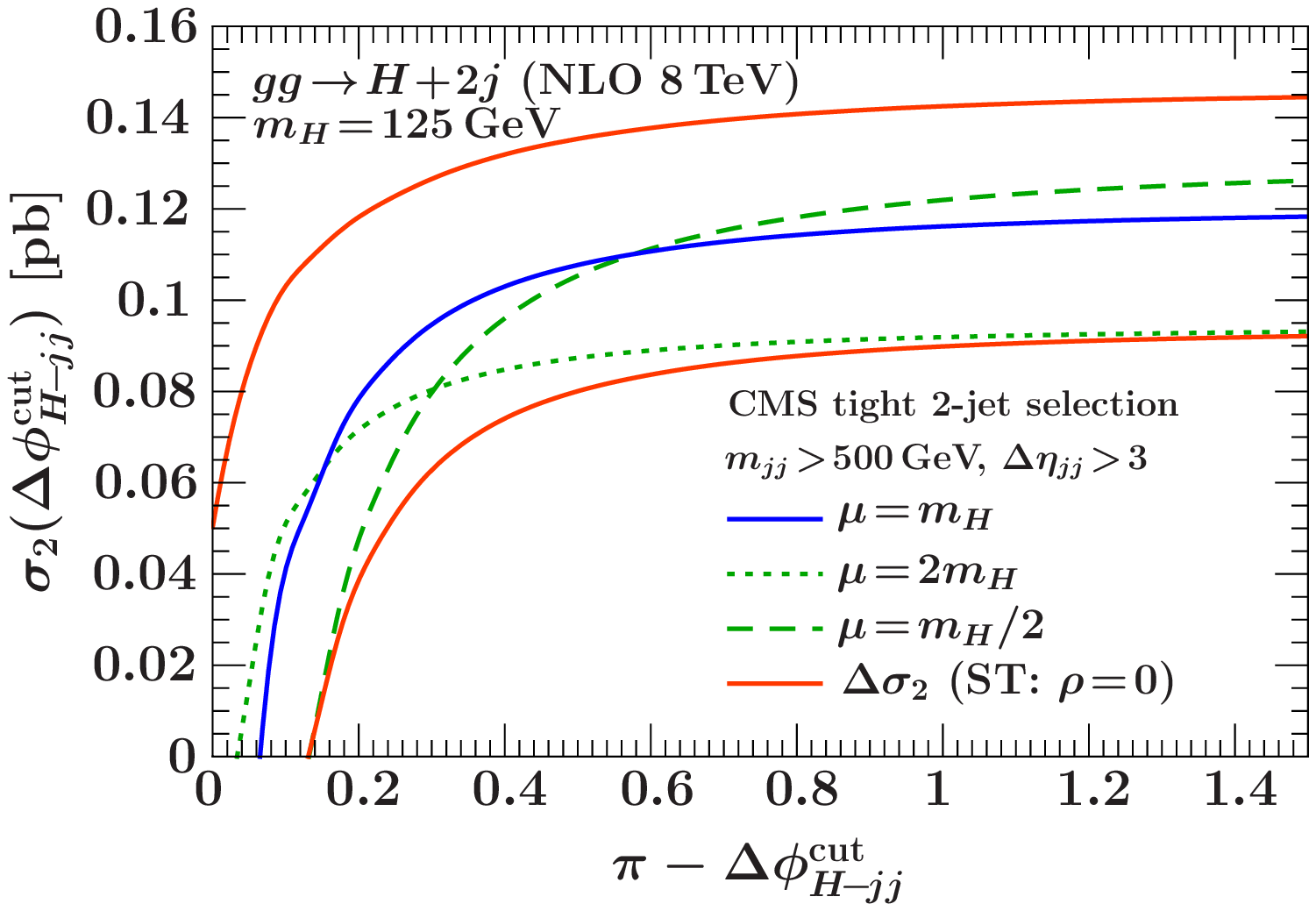}\label{fig:cmstightphi}}%
\caption{Exclusive $pp\to H + 2$ jet cross section via ggF at NLO for as function of $\pT^\cut$ (left panels) and $\pi-\dphi^\cut$ (right panels) for both ATLAS and CMS VBF selections.}
\label{fig:ptphiresults}
\end{figure*}

\section{Results}
\label{sec:results}

\begin{table*}[t!]
\begin{tabular}{c|c|cc|c}
\hline\hline
Selection & $\sigma\, [\mathrm{pb}]$  & \multicolumn{2}{|c|}{ {Direct scale variation}} & {Combined incl. uncertainties} \\
 & $\mu=m_H$ & $\mu=2m_H$ & $\mu=m_H/2$ &  ST ($\rho = 0$) \\
\hline\hline
\multicolumn{5}{c}{\textbf{ATLAS}}\\
\hline
$\sigma_{\ge 2}$ & $0.21$ & & & $\pm 21\%$   \\
$\sigma_2(\pT < 30\GeV)$ & $0.15$ & $-8 \%$ & $-29 \% $ & $\pm 44 \%$  \\
$\sigma_2(\dphi > 2.6)$ & $0.19$ & $-17 \%$ & $-4 \% $ & $\pm 26\% $ \\
$\sigma_2(\pT < 30\GeV\!,\, \dphi > 2.6)$ & $0.14$ & $ -5 \%$ & $-45 \%$ & $ \pm 56 \% $ \\
\hline\hline
\multicolumn{5}{c}{\textbf{CMS loose}}\\
\hline
$\sigma_{\ge 2} $ & $0.41$ & & & $\pm 21 \%$ \\
$\sigma_2(\pT < 30\GeV)$ & $0.35$ & $-18 \%$ & $0 \%$  & $ \pm 28 \%$  \\
$\sigma_2(\dphi > 2.6)$ & $0.39$ & $-20\%$ & $+9 \%$ &  $\pm 24 \% $ \\
$\sigma_2(\pT < 30\GeV\!,\, \dphi > 2.6)$ & $0.34$ & $-16 \%$ & $-4 \%$ & $\pm 31 \% $  \\
\hline\hline
\multicolumn{5}{c}{\textbf{CMS tight}}\\
\hline
$\sigma_{\ge 2} $ & $0.12$ & & & $\pm 21 \%$ \\
$\sigma_2(\pT < 30\GeV)$ & $0.08$ & $-8 \%$ & $-35 \%$  & $ \pm 49 \%$  \\
$\sigma_2(\dphi > 2.6)$ & $0.10$ & $-19\%$ & $-1 \%$ &  $\pm 26 \% $ \\
$\sigma_2(\pT < 30\GeV\!,\, \dphi > 2.6)$ & $0.07$ & $-7 \%$ & $-46 \%$ & $\pm 53 \% $  \\
\hline\hline
\end{tabular}
\caption{Perturbative uncertainties at NLO in the exclusive $pp\to H + 2$ jet cross section via gluon fusion for cuts on $\pT$ and $\dphi$ for both ATLAS and CMS VBF selections. }
\label{tab:numbers}
\end{table*}

In this section, we present our results for the exclusive $pp\to H+2$ jet cross section via ggF at NLO, taking ST with $\rho = 0$ as our method of choice to estimate the perturbative uncertainties. All our inputs are summarized at the beginning of \sec{application}. The ATLAS, CMS loose, and CMS tight VBF selection cuts we apply are summarized in Table~\ref{tab:cuts}.

\subsection{\boldmath $gg\to H + 2$ Jets Cross Section }
\label{subsec:results}

\subsubsection{$\pT$ and $\dphi$}

In \fig{ptphiresults} we plot the result for the exclusive 2-jet cross section as a function of $\pT^\cut$ and $\dphi^\cut$ for the ATLAS, CMS loose, and CMS tight VBF selections. In all our cross section plots the solid blue central line shows the central-value prediction obtained from $\mu = m_H$, while the outer orange solid lines show our uncertainty estimate. For reference, the green dashed and dotted curves show the direct scale variation for $\mu =m_H/2$ and $\mu=2m_H$, respectively.

The overall picture is very similar for all three VBF selections and both binning variables. For large values of $\pT^\cut$ or $\pi -\dphi^\cut$, the cross section $\sigma_{\ge 3}$ that is cut away becomes small and so the effect of $\Delta^\cut$ is negligible. In this limit the uncertainties reproduce those in the inclusive 2-jet cross section, which here are determined by the $\mu = 2m_H$ variation (cf. \subsec{inclscales}). On the other hand, in the transition region, once the exclusive cut starts to impact the cross section, the direct scale variations cannot be used any longer to estimate uncertainties, which is exhibited by the crossing of the lines. As explained in detail in the previous two sections, the reason is that the direct scale variation only gives an estimate of the yield uncertainties, which effectively assumes the scale variations in the inclusive cross sections to be $100\%$ correlated (corresponding to $\rho=1$). At the same time it neglects the migration uncertainty in the binning, which becomes important as the exclusive cut gets tighter. In the ST procedure, this effect is taken into account explicitly, which thus gives more robust uncertainties for all values of $\pT^\cut$ or $\dphi^\cut$.

In Table~\ref{tab:numbers} we quote results for the cross sections and their percentage uncertainties for specific cuts. For $\dphi$ we use the current experimental value $\dphi > 2.6$. Compared to the $21\%$ in the inclusive 2-jet cross section with VBF cuts ($\sigma_{\geq 2}$), we see a moderate increase in the uncertainty in $\sigma_2(\dphi > 2.6)$ to $26\%$ for ATLAS and CMS tight, and $24\%$ for CMS loose. For $\pT$ we use a representative value of $\pT < 30\GeV$, for which the uncertainties increase substantially to $44\%$ and $49\%$ for ATLAS and CMS tight, and moderately to $28\%$ for CMS loose. Note that for a fixed cut the uncertainties increase with a tighter VBF selection. This is also clearly visible in \fig{ptphiresults}, where the region where the cross section drops and the uncertainties grow large moves to larger values of $\pT$ or $\pi-\dphi$, going from CMS loose to ATLAS to CMS tight. We will come back to this in \subsec{ggFVBF}.

\subsubsection{Combination of Exclusive Cuts}

\begin{figure*}[t!]
\includegraphics[width=1.03\columnwidth]{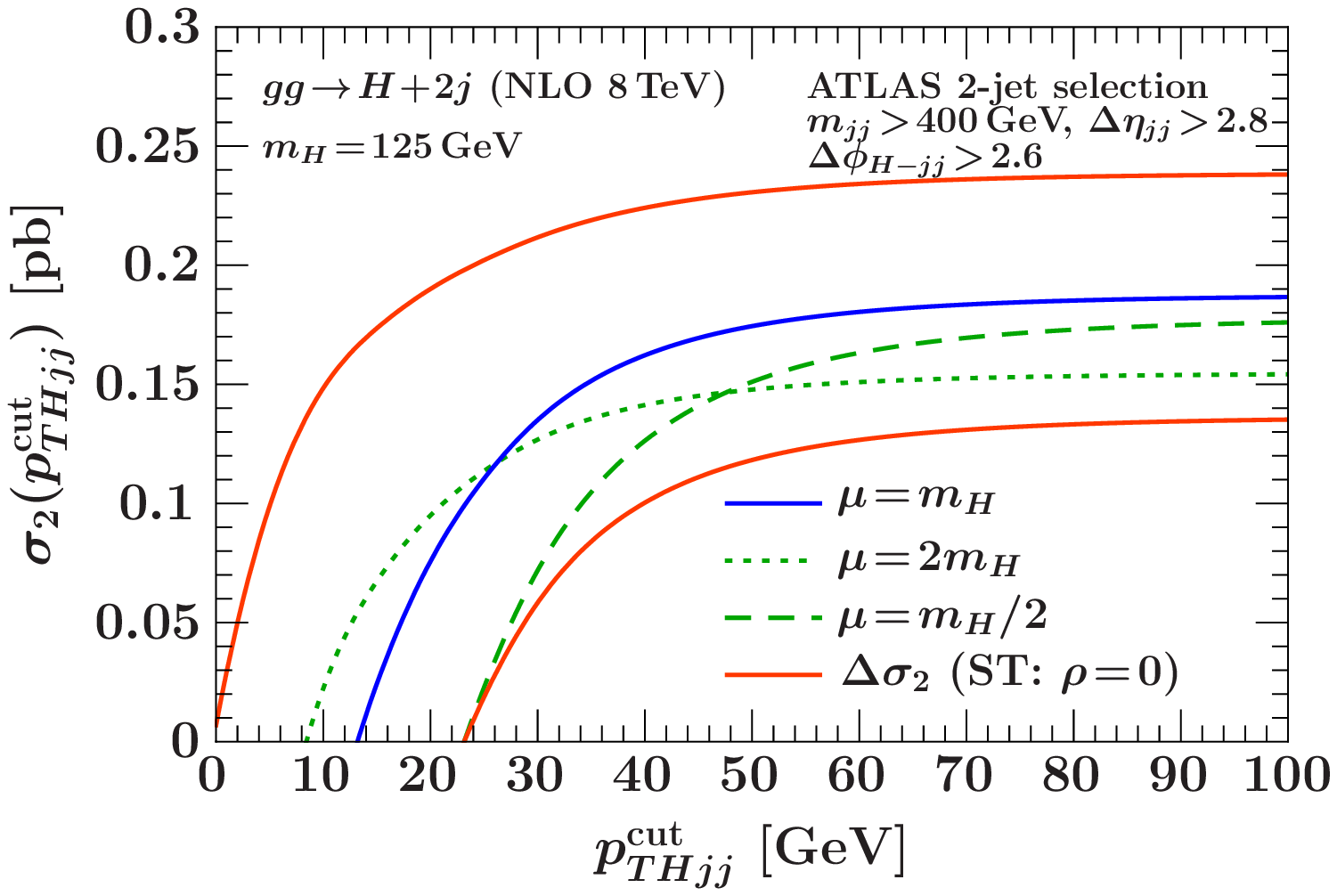}%
\hfill%
\includegraphics[width=\columnwidth]{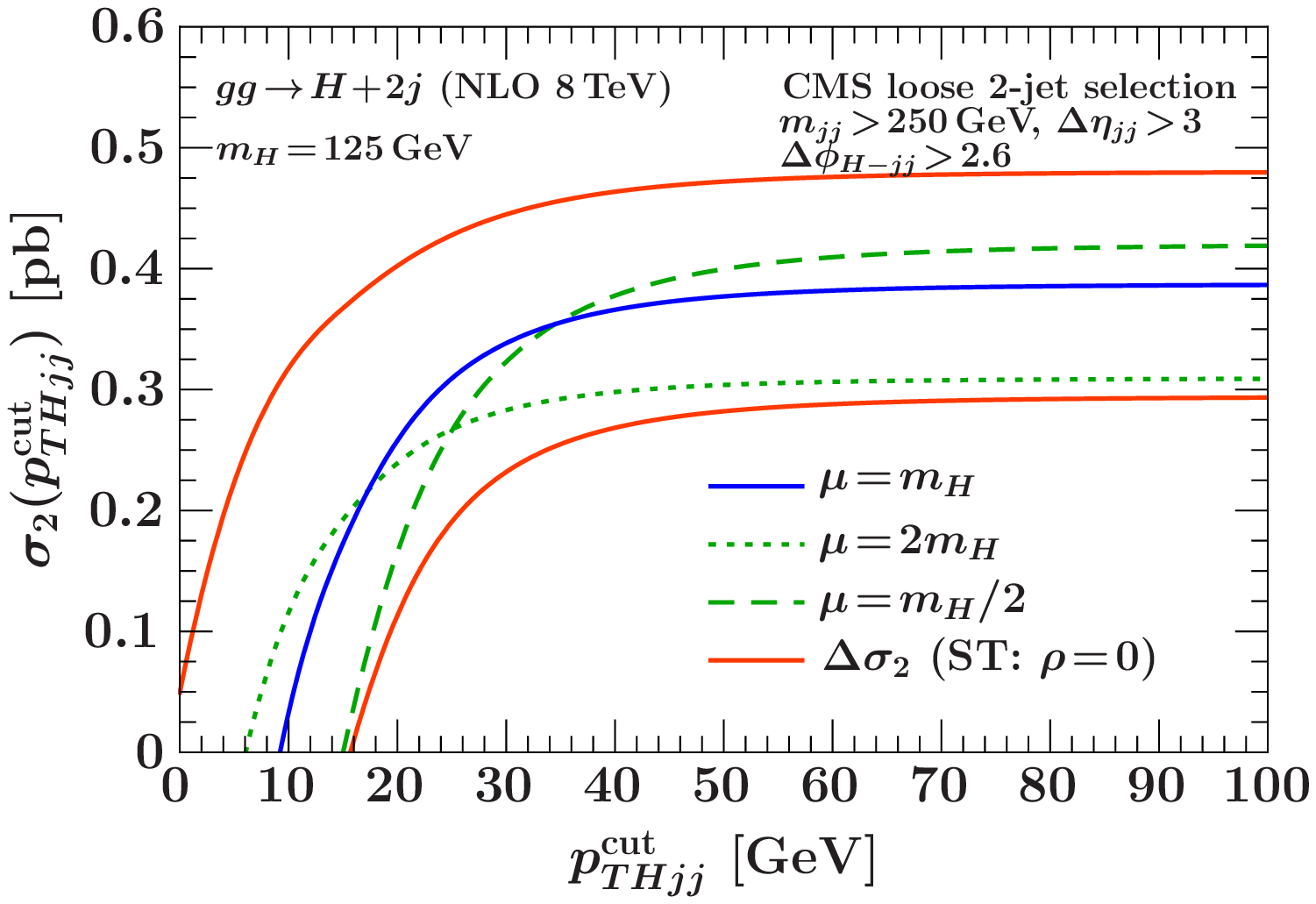}%
\vspace{-2ex}
\caption{Exclusive 2-jet cross section as a function of $\pT^\cut$ with an additional cut $\dphi > 2.6$ using the ATLAS (left panel) and CMS loose (right panel) VBF selections.}
\vspace{-2ex}
\label{fig:pTdphi}
\end{figure*}

As the cases of $\dphi$ and $\pT$ already show, one has to be careful when cutting on variables which effectively force the kinematics in the exclusive 2-jet region and induce large logarithms in the perturbative series. Whether implementing a cut-based approach or in multivariate analysis, it is important to take into account the uncertainties induced by the exclusive restriction. As an illustration of the application of the ST method to a more general case, we now consider the case where we combine cuts on both $\pT$ and $\dphi$.

Specifically, we study the exclusive 2-jet cross section as a function of $\pT^\cut$ with an additional constraint that we select only events which already have $\dphi > 2.6$. Following \eq{sigmaNgeneral}, the corresponding exclusive 2-jet cross section can be expressed as
\begin{align}
&\sigma_{2}(\dphi > 2.6,\, \pT< \pT^\cut)
\\\nn & \qquad
= \sigma_{\geq 2} - \sigma_{\geq 3}(\dphi < 2.6\,\,\text{or}\,\, \pT > \pT^\cut)
\,.\end{align}
Taking $\rho = 0$ for simplicity, the corresponding exclusive uncertainty is now given in terms of the uncertainties obtained by scale variation in the inclusive cross sections as
\begin{align} \label{eq:Delta2combined}
&\Delta_2^2(\dphi > 2.6,\, \pT < \pT^\cut)
\\\nn & \quad
= \Delta_{\geq 2}^{\mu\, 2} + \Delta^{\mu\,2}_{\geq 3}(\dphi < 2.6\,\,\text{or}\,\, \pT > \pT^\cut)
\,.\end{align}

In \fig{pTdphi}, we show $\sigma_2$ as a function of the $\pT^\cut$ with fixed $\dphi > 2.6$ for the ATLAS and CMS loose VBF selections. As before, the cross section for $\mu = m_H$ is the central solid blue curve and the green dashed and dotted curves show the result of direct scale variation by a factor of two, while the outer solid orange lines show the uncertainties obtained from \eq{Delta2combined}. As shown in Table~\ref{tab:numbers}, for $\pT^\cut = 30 \GeV$ we now get $56\%$, $31\%$, and $53\%$ uncertainty for ATLAS, CMS loose, and CMS tight, which is slightly increased compared to not having the additional cut on $\dphi$. For large values of $\pT^\cut$ the uncertainties in \fig{pTdphi} correctly reproduce the exclusive uncertainties for $\Delta_2(\dphi> 2.6)$ without the cut on $\pT$ [see \figs{atlasphi}{cmsloosephi}].

\vspace{-2ex}
\subsection{Uncertainties in ggF-VBF Separation}
\vspace{-1ex}
\label{subsec:ggFVBF}

\begin{figure*}[t!]
\subfigure[ATLAS VBF selection]{\includegraphics[width=1.03\columnwidth]{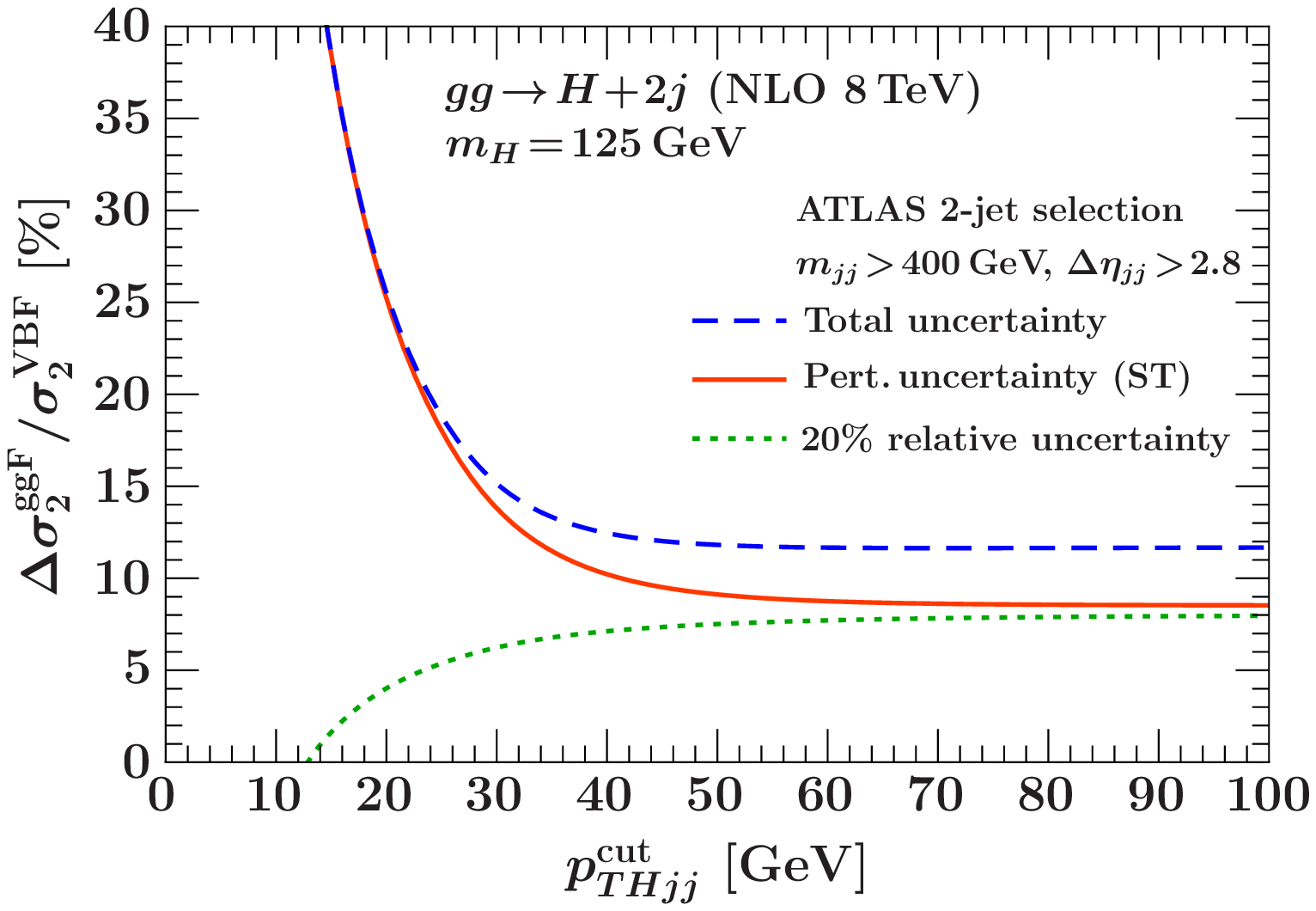}\label{fig:atunc}}%
\hfill%
\subfigure[ATLAS VBF selection]{\includegraphics[width=\columnwidth]{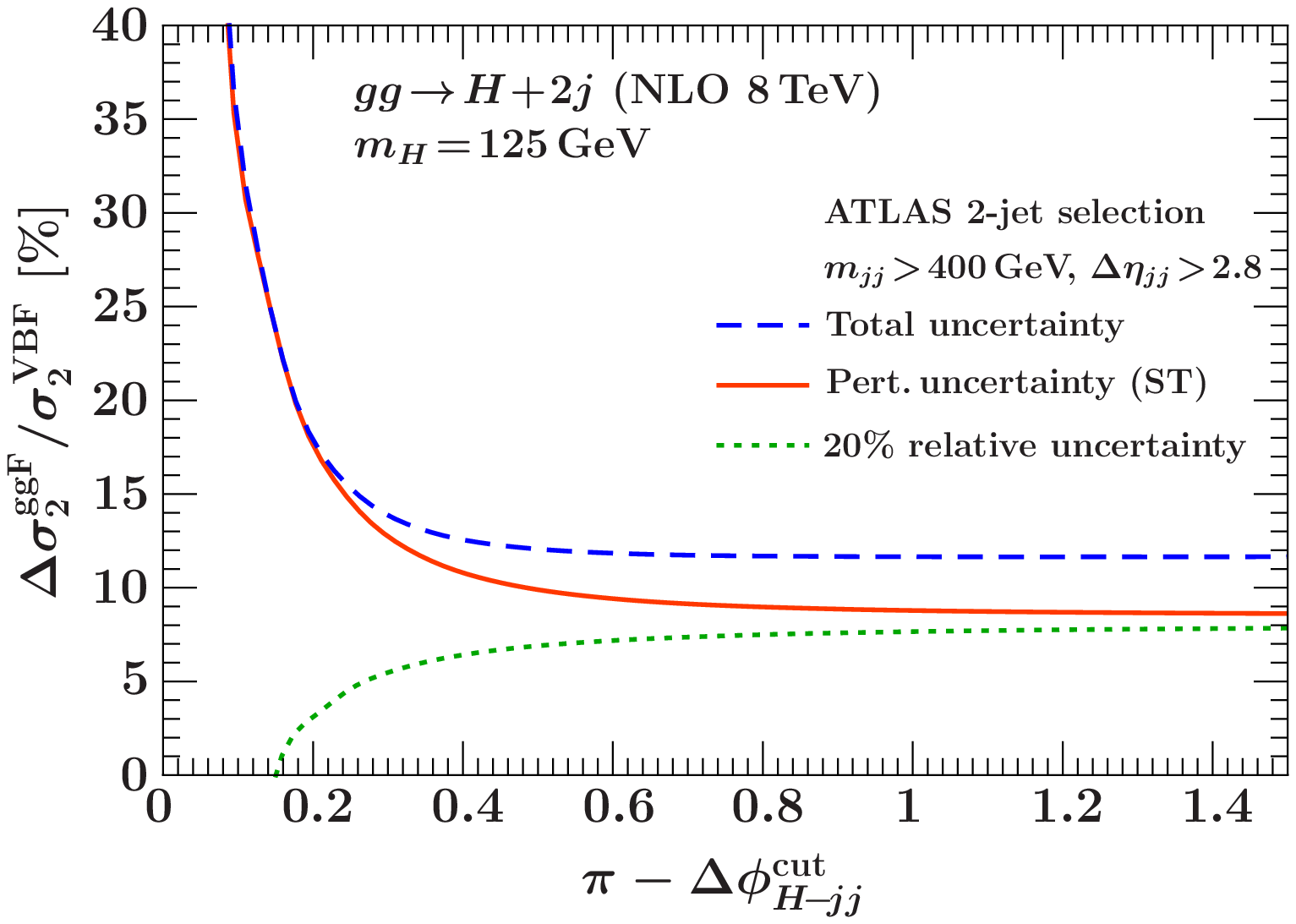}\label{fig:atuncphi}}%
\\
\subfigure[CMS loose VBF selection]{\includegraphics[width=1.03\columnwidth]{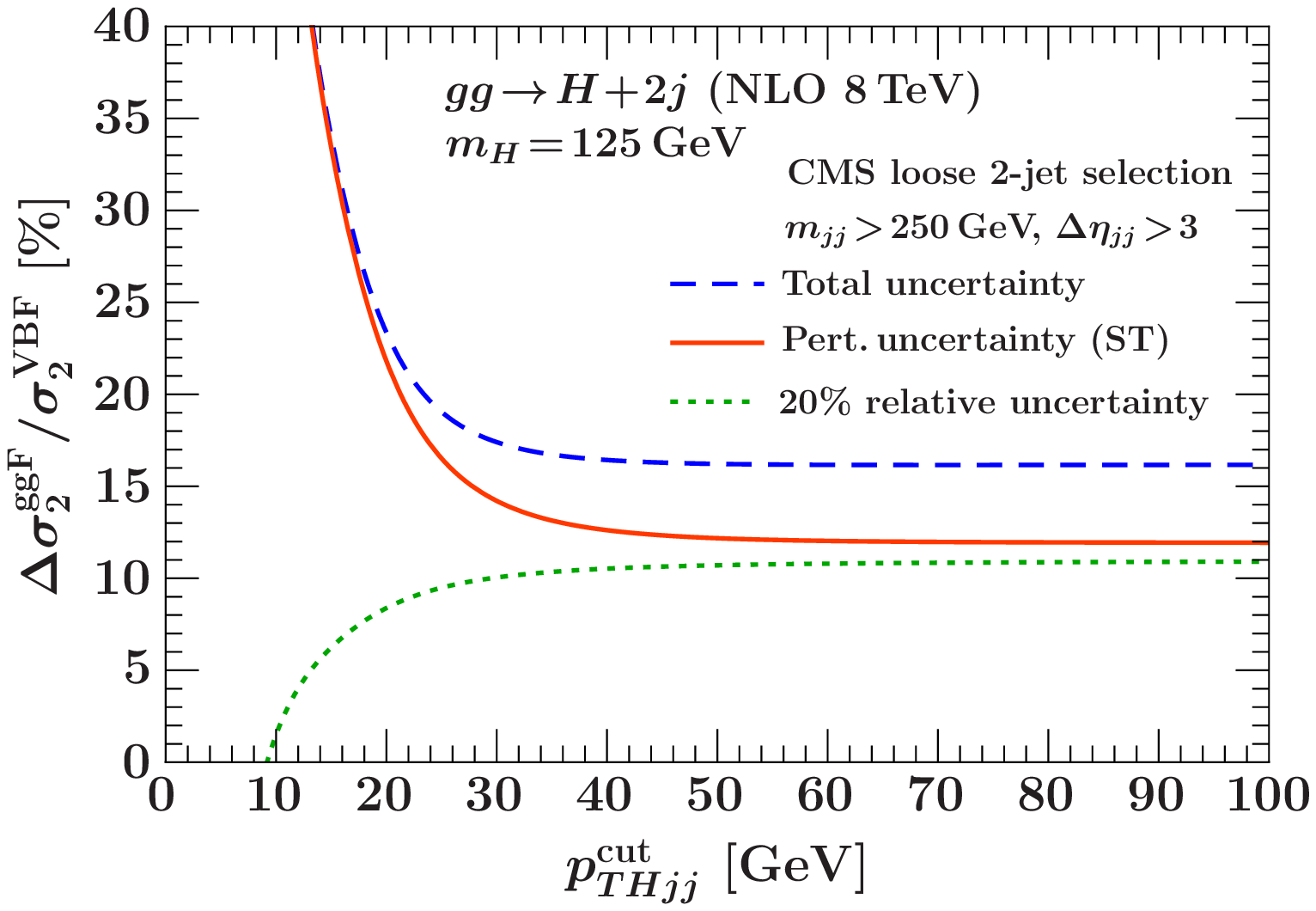}\label{fig:cmsunc}}%
\hfill%
\subfigure[CMS loose VBF selection]{\includegraphics[width=\columnwidth]{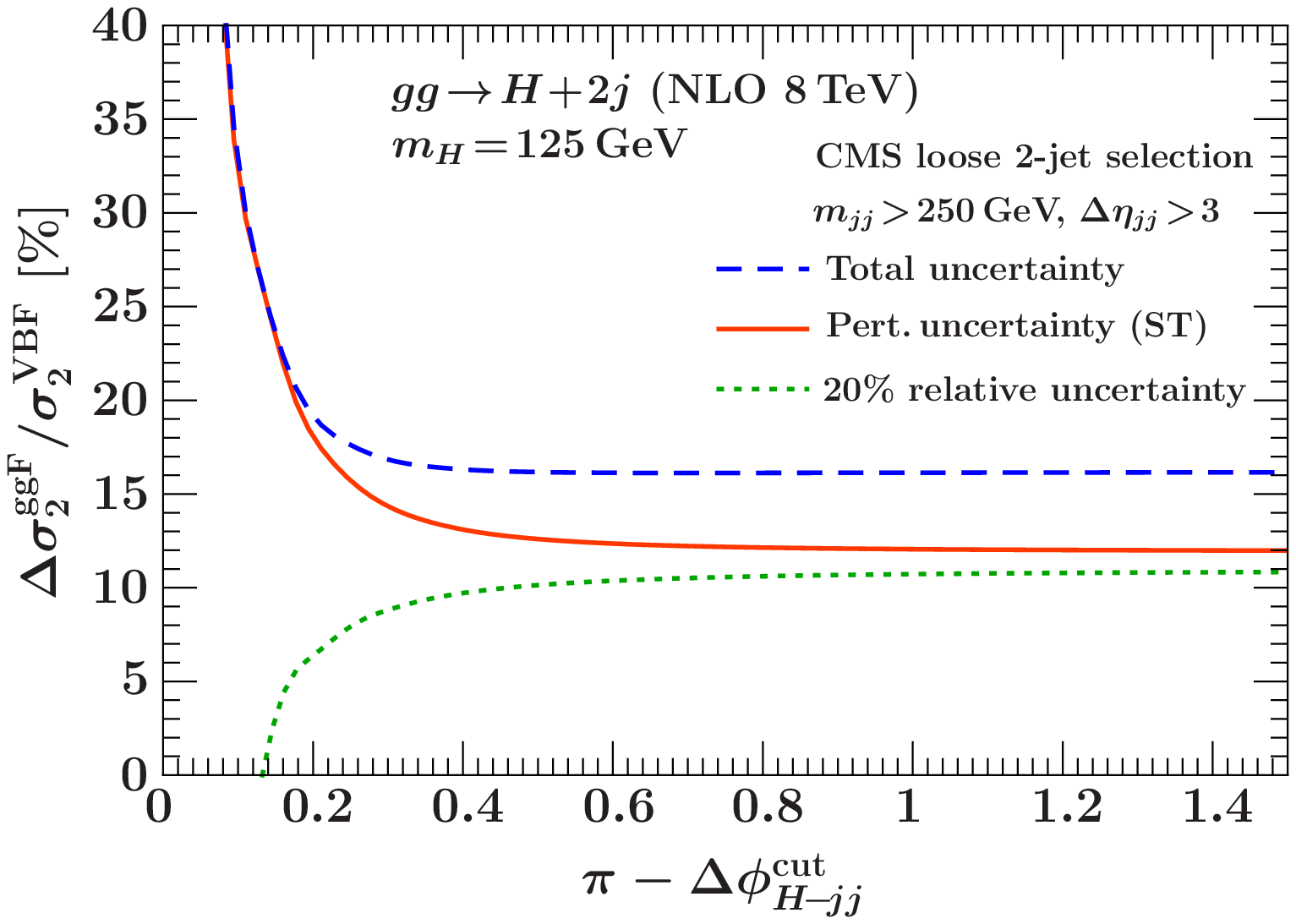}\label{fig:cmsuncphi}}%
\caption{Theoretical uncertainties of the ggF contribution relative to the VBF cross section as function of $\pT^\cut$ (left panels) and $\dphi^\cut$ (right panels) for the ATLAS VBF selection (top panels) and CMS loose VBF selection (bottom panels). The solid orange lines show the perturbative uncertainties in $\sigma_2^\mathrm{ggF}$, the green dotted lines a flat $20\%$ parametric uncertainty in $\sigma_2^\mathrm{ggF}$, and the dashed blue lines both contributions added in quadrature.}
\label{fig:ggFVBFunc}
\end{figure*}

The VBF production process is characterized by two forward jets with large rapidity separation and large dijet invariant mass. The VBF selection cuts used by the ATLAS and CMS experiments enhance the VBF contribution, but a significant $\sim 25\%$ ggF contribution remains. Since the VBF cross section is known rather precisely, an important source of theoretical uncertainty in the extraction of the VBF signal is the large perturbative uncertainty in the ggF contribution. After subtracting the non-Higgs backgrounds (which are of course another source of uncertainty), the measured cross section for Higgs production after implementing the VBF selection is given by
\begin{equation}
\sigma_2^\mathrm{measured}(\dphi^\cut) = \sigma_2^\mathrm{VBF}(\dphi^\cut) + \sigma_2^\mathrm{ggF}(\dphi^\cut)
\,.\end{equation}
For the purpose of extracting the VBF cross section, we effectively have to subtract the theory prediction for $\sigma_2^\mathrm{ggF}(\dphi^\cut)$ from $\sigma_2^\mathrm{measured}(\dphi^\cut)$. Therefore, the relevant figure of merit is $\Delta\sigma_2^\mathrm{ggF}(\dphi^\cut)/\sigma_2^\mathrm{VBF}(\dphi^\cut)$, i.e., the theory uncertainty in $\sigma_2^\mathrm{ggF}$ measured relative to the expected VBF cross section, $\sigma_2^\mathrm{VBF}$.

\begin{figure*}[t!]
\includegraphics[width=\columnwidth]{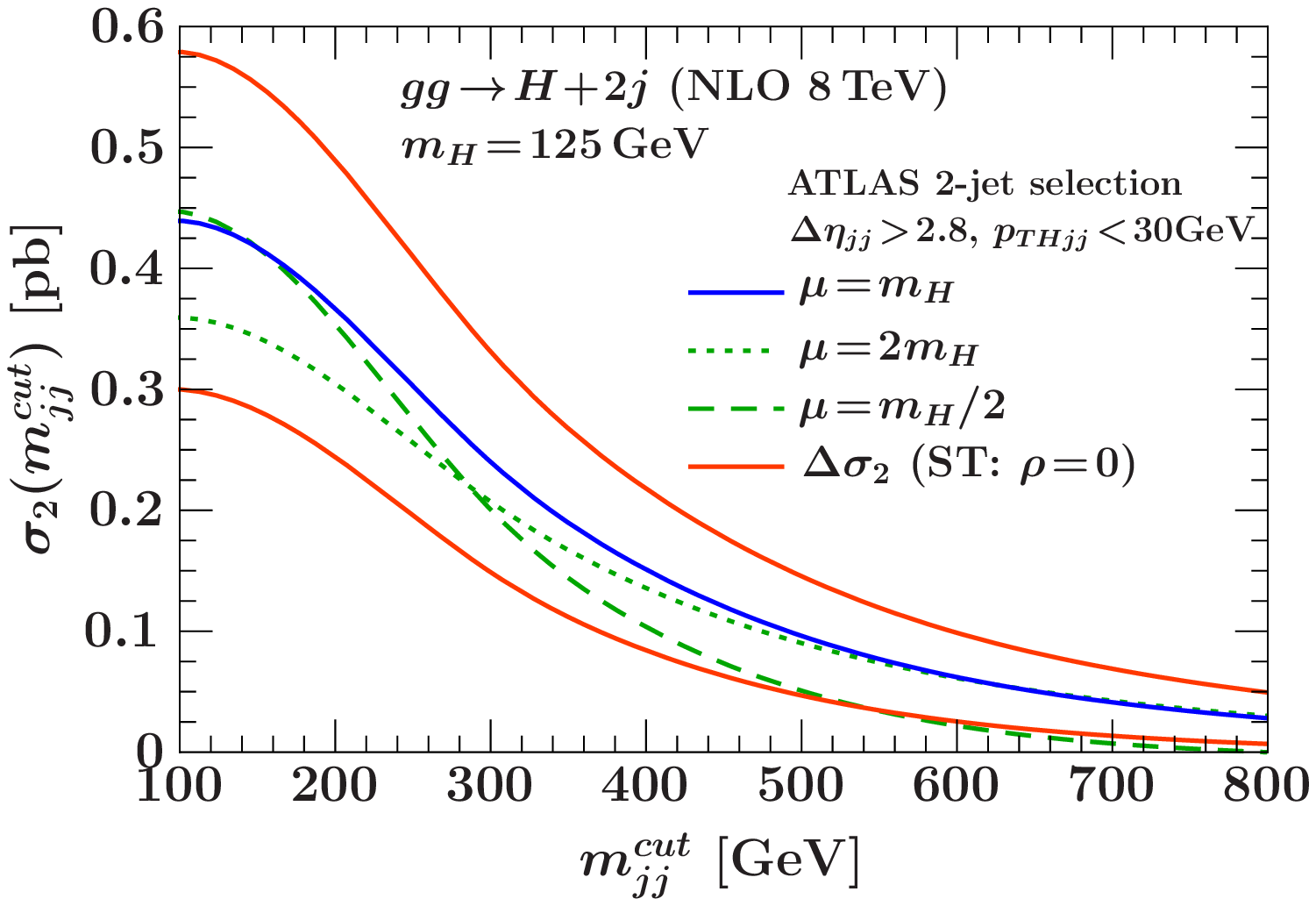}%
\hfill%
\includegraphics[width=\columnwidth]{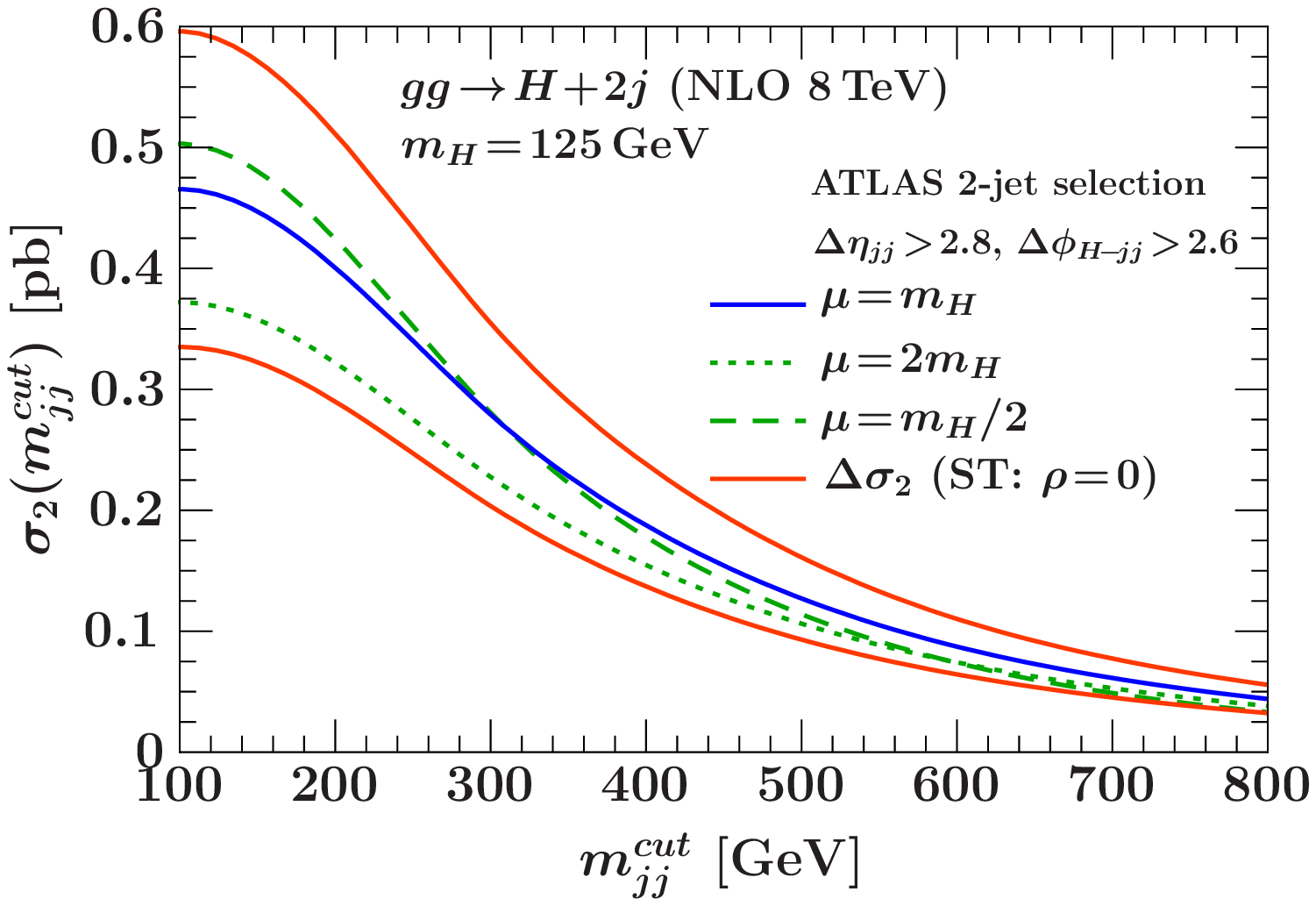}%
\caption{Exclusive 2-jet cross section over a range of $m_{jj}^\cut$ for fixed $\pT<30\GeV$ (left panel) and fixed $\dphi > 2.6$ (right panel) for the ATLAS VBF selection.}
\label{fig:atmjjuncpt}
\end{figure*}

\begin{figure*}[t!]
\includegraphics[width=\columnwidth]{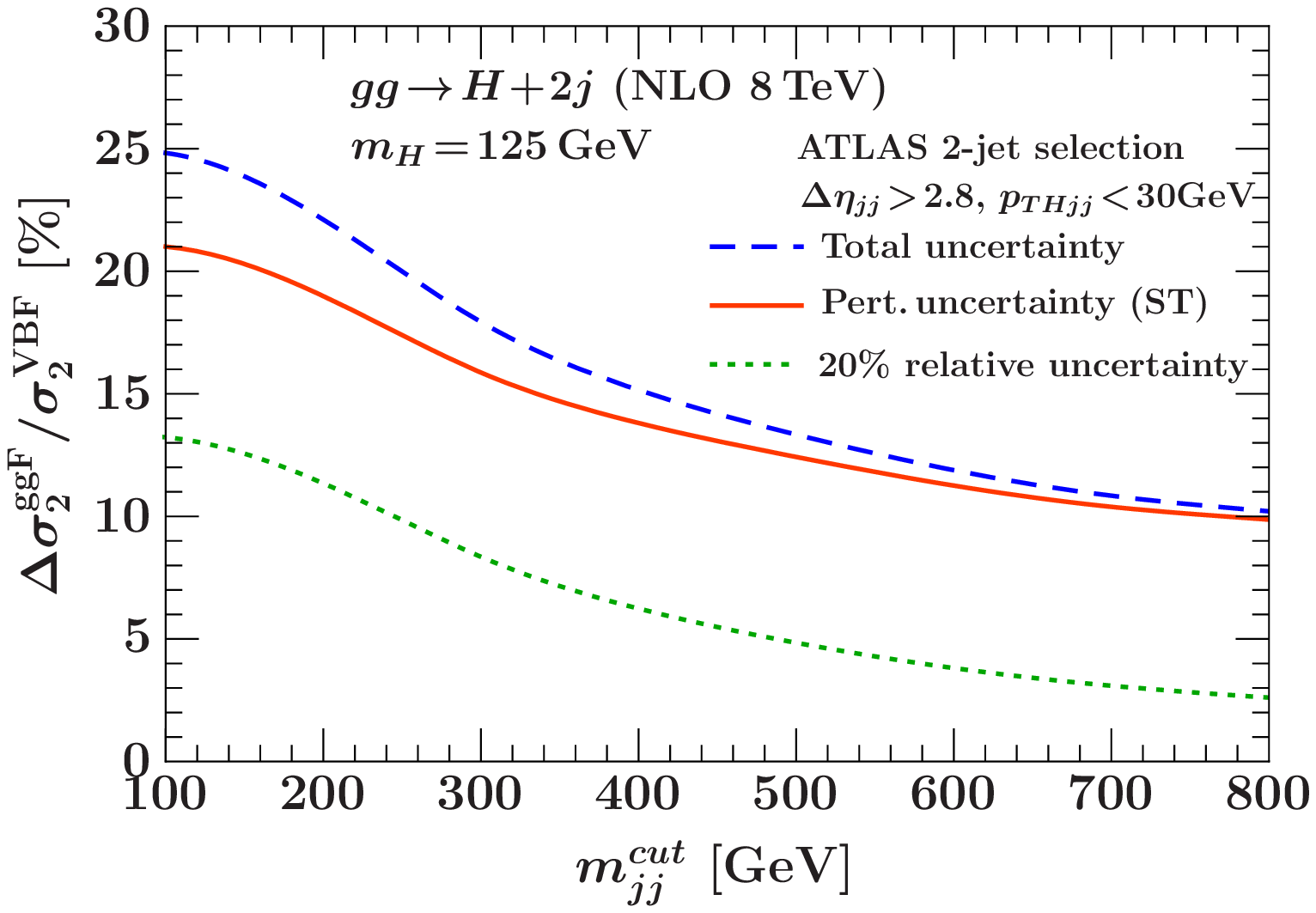}%
\hfill%
\includegraphics[width=\columnwidth]{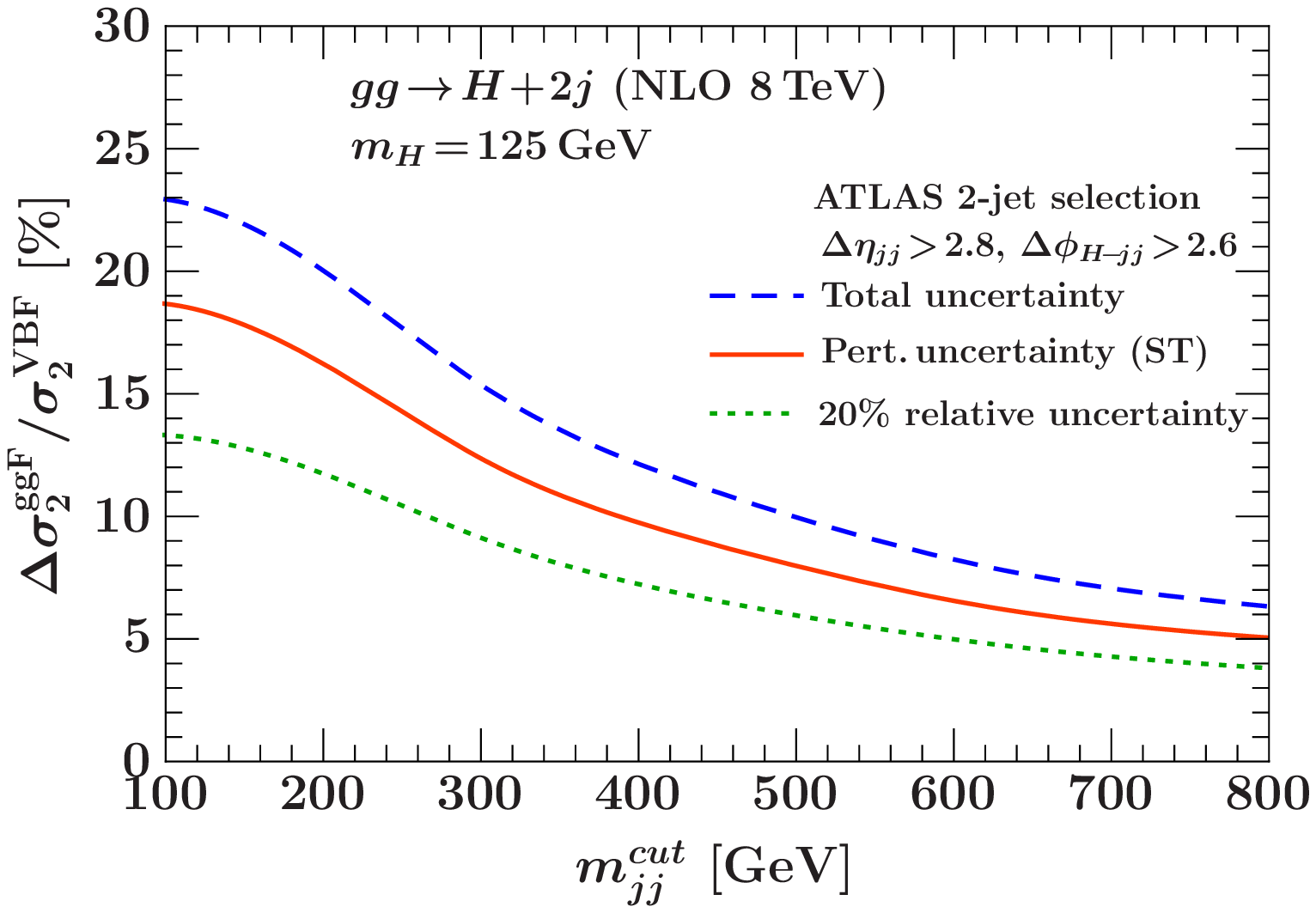}%
\caption{Perturbative uncertainties of the ggF contribution relative to the VBF cross section over a range of $m_{jj}^\cut$ for fixed $\pT<30\GeV$ (left panel) and fixed $\dphi > 2.6$ (right panel) for the ATLAS VBF selection.}
\label{fig:atvbfggfphi}
\end{figure*}

In \fig{ggFVBFunc} we show the ggF uncertainty relative to the VBF signal cross section over a range of $\pT^\cut$ and $\dphi^\cut$ using the ATLAS and CMS loose VBF selections. In these plots, the solid orange curve shows our results for the NLO perturbative uncertainties (corresponding to the orange lines in \fig{ptphiresults}). For comparison, the green dotted curve shows a fixed $20\%$ uncertainty in the ggF cross section, i.e., taking $\Delta\sigma_2^\mathrm{ggF} = 0.2\, \sigma_2^\mathrm{ggF}$, which for example could be due to PDF and $\alpha_s$ parametric uncertainties. Hence, the green dotted lines effectively track the size of the ggF cross section relative to the VBF cross section (multiplied by $0.2$). In the dashed blue lines, both uncertainty contributions are added in quadrature.

In the region of low $\pT^\cut$ or $\pi-\dphi^\cut$, the relative uncertainty coming from the ggF contribution quickly increases below $\pT\lesssim 30\GeV$ and $\pi-\dphi \lesssim 0.4$. This is despite the fact that the relative ggF cross section quickly decreases there, as can be inferred from the decrease in the dotted green lines. In this region, the total uncertainty shown by the blue dashed curve becomes completely dominated by the perturbative ggF uncertainty. Hence, one should be careful when implementing and optimizing either indirect restrictions on additional radiation, like $\dphi$, or explicit $p_T$-vetoes like $\pT$, since the gain in sensitivity in the Higgs signal from reduced non-Higgs backgrounds must be weighed against the increased theoretical uncertainty in separating the ggF and VBF contributions.

We already saw in \subsec{results} that the perturbative uncertainties in the exclusive 2-jet cross section also depend on the chosen VBF cuts and increase with a higher cut on the dijet invariant mass, $m_{jj}$. The reason for this effect is that at higher $m_{jj}$ the effective hard scale in the process is also pushed higher causing the logarithmic corrections at a given value of $\pT^\cut$ to increase. This is seen explicitly in \fig{atmjjuncpt}, which shows the exclusive 2-jet cross section over a range of $m_{jj}^\cut$ using the ATLAS VBF selection for a fixed cut $\pT < 30\GeV$ or $\dphi > 2.6$, where the curves have the same meaning as in \figs{ptphiresults}{pTdphi}. As expected, with a cut on $\pT<30\GeV$, we see that the relative uncertainty in the ggF cross section grows for larger $m_{jj}$ values, and reaches almost $100\%$ for $m_{jj} \gtrsim 800\GeV$. Note however that for such large $m_{jj}$ cuts one might have to reevaluate whether $\mu = m_H$ is still an appropriate scale choice for this process. With a cut on $\dphi > 2.6$, the relative uncertainty in the ggF cross section stays roughly constant for larger $m_{jj}$ presumably because this cut is somewhat milder, which we also saw in the results in Table~\ref{tab:numbers}.

In \fig{atvbfggfphi} we show the ggF uncertainty relative to the VBF cross section analogous to \fig{ggFVBFunc}. We can clearly see that in this case tightening the cut on $m_{jj}$ does improve the separation of the ggF and VBF contributions, as the perturbative ggF uncertainty relative to the VBF cross section, shown by the orange curves, decreases. In this case, the overall reduction of the ggF contamination relative to the VBF cross section is stronger than the increase in the perturbative uncertainties of the ggF contribution.

\section{Conclusions}
\label{sec:conclusions}

In order to enhance the VBF signal over non-Higgs backgrounds as well as the ggF contribution, the typical VBF selection cuts used by the ATLAS and CMS experiments include either indirect or direct restrictions on additional emissions. Such restrictions constitute a nontrivial jet binning, where the inclusive 2-jet cross section is effectively divided into an exclusive 2-jet bin and a remaining inclusive 3-jet bin.

With such a jet binning one has to account for two sources of perturbative uncertainties. In addition to the absolute yield uncertainty which is correlated between the jet bins, there is also a migration uncertainty which is anticorrelated and drops out in the sum of the bins. This migration uncertainty is associated with the perturbative uncertainty in the logarithmic series that is introduced by the exclusive binning cut. As the binning cut becomes tighter, the logarithms grow large and eventually lead to a breakdown of fixed-order perturbation theory, at which point a logarithmic resummation becomes necessary.

In practice, the experimentally relevant region typically lies inside the transition region between the fully inclusive region (no binning) and the extreme exclusive region (very tight binning). In this region, fixed-order perturbation theory can still be applied. However, since the logarithms are already sizeable, one has to explicitly take into account the migration uncertainty. This can be achieved using the ST method.

We studied in detail the application of the ST method for $pp \to H+2$ jets via ggF, including its generalization and validation against alternative prescriptions. We find that the perturbative uncertainties are very sensitive to the exclusive cut and can quickly become sizeable. While applying a strong restriction on additional emissions is expected to increase the sensitivity to the VBF signal, it is not necessarily beneficial for distinguishing the VBF and ggF production modes because of the quickly increasing ggF uncertainties. Hence, it would be important to include the perturbative uncertainties as a function of the binning cut when optimizing the experimental selections.

\vspace{-1ex}
\begin{acknowledgments}
We thank Florian Bernlochner and Dag Gillberg for helpful discussions and comments
on the manuscript. This work was supported by the DFG Emmy-Noether Grant No. TA 867/1-1.
\end{acknowledgments}

\bibliographystyle{physrev4}
\bibliography{vbf}

\end{document}